\documentclass[preprint,journal]{vgtc}            % preprint (journal style)

\onlineid{0}

\vgtccategory{Research}

\title{The Frustrometer: Detecting User Frustration in Data Visualization Tasks using Biomarkers and Interaction Patterns}

\author{%
    \authororcid{Johannes Ellemose}{0009-0006-9676-6204},
    \authororcid{Sophia Wanner}{0009-0001-1331-4776},
    \authororcid{Djordje Slijepčević}{0000-0002-2295-7466},
    \authororcid{Laura Cesar}{},\\
    \authororcid{Vanessa Leung}{0000-0003-0388-9864},
    \authororcid{Wolfgang Aigner}{0000-0001-5762-1869}, and 
    \authororcid{Niklas Elmqvist}{0000-0001-5805-5301}
}

\authorfooter{
  \item Johannes Ellemose and Niklas Elmqvist are with the Department of Computer Science at Aarhus University, Aarhus, Denmark.
        E-mail: \{ellemose, elm\}@cs.au.dk.

  \item Sophia Wanner, Djordje Slijepčević, Laura Cesar, Vanessa Leung and Wolfgang Aigner are with the University of Applied Sciences St. Pölten, St. Pölten, Austria.
        E-mail: \{sophia.wanner, djordje.slijepcevic, laura.cesar, vanessa.leung, wolfgang.aigner\}@ustp.at.
}

\abstract{%
    Visualization research has largely solved \textit{how} to help a stuck or frustrated user---through interactive onboarding, contextual help, and active guidance.
    The unsolved problem is \textit{when}: trigger help too eagerly and you break the user's train of thought; wait too long and they have already gone astray.
    We present the \textsc{Frustrometer}, a series of experiments to predict user stuckness and frustration by fusing physiological and interaction signals.
    The Frustrometer consists of 
    a convolutional neural network classifier, that in real-time estimates whether user are stuck in their task or not. 
    We collected data from a controlled study where 14 participants performed analytical tasks on two interactive visualization dashboards while we captured eye movement, pupil dilation, galvanic skin response, heart-rate, head orientation, mouse dynamics, and keyboard events. 
    In addition participants assessed their own performance, while we annotated when during the tasks the participants were stuck. 
    Our results reveal that autonomous physiological responses such as heart-rate and galvanic skin response provide limited insights into the frustration level of the user. 
    Similarly, head orientations are not easily correlated with the frustrations felt by the user during visual analysis tasks. 
    Mouse movements and gaze data conversely carry the majority of predictive signal, with mouse movements alone having a strong correlation for some participants, suggesting that lightweight instrumentation may suffice for real-time frustration detection. 
    We end the paper by discussing how these findings can inform the design of adaptive guidance systems for complex visualization tasks that takes a multimodal approach to frustration and stuckness detection. 
}


\teaser{
  \centering
  \includegraphics[width=\linewidth, alt={The Frustrometer workflow.}]{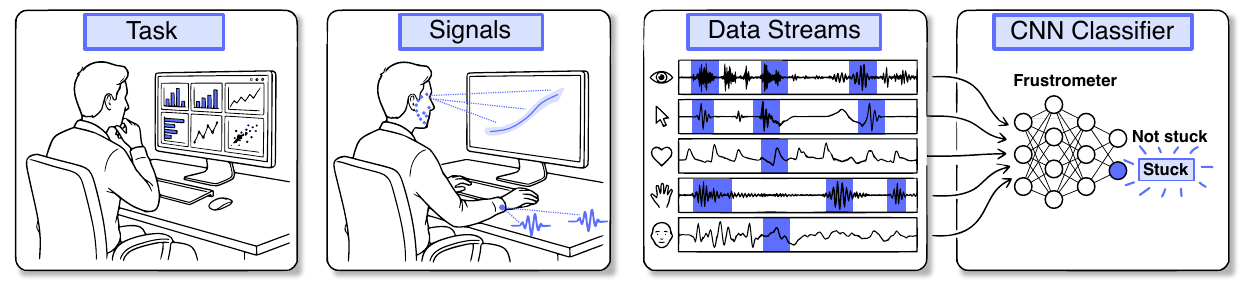}
  \caption{%
    \textbf{The Frustrometer.}
    A user performs analytical tasks on a visualization dashboard (Panel 1) while physiological and interaction signals---eye gaze, mouse dynamics, heart rate, galvanic skin response, and facial expression---are captured simultaneously (Panel 2).
    These multimodal data streams are temporally aligned and segmented, with stuckness episodes annotated as ground truth (Panel 3), then fed into a convolutional neural network that classifies whether the user is stuck or not---a precursor to frustration (Panel 4).
  }
  \label{fig:teaser}
}

\graphicspath{{figs/}{figures/}{pictures/}{images/}{./}} % where to search for the images

\usepackage{tabu}                      % only used for the table example
\usepackage{booktabs}                  % only used for the table example
\usepackage{lipsum}                    % used to generate placeholder text
\usepackage{fontawesome5}
\usepackage{mathptmx}                  % use matching math font

\usepackage[table,xcdraw]{xcolor}
\usepackage{placeins}

\input{customtodos}

\begin{document}

\firstsection{Introduction}

\maketitle

%% ---------------------------------------------------------------------
%% Content
%% ---------------------------------------------------------------------

% %% TODO: 
% - on~track vs on-track DONE 
% - biomarkers vs physiological data vs biometric data vs body signals ==> which one to choose?? BE CONSISTENT DOE 
% - introduce impasse as cognitive psych's fancy term?? 
%     - "proportional time in impasse" or "normalized impasse duration" rather than % of time stuck ==> It sounds fancier
%         - update in abstract, introduction, RW, figures, discussion, stats analysis etc. 
% - streamline frustration vs stuckness ==> we are detecting stuckness, which is a precursor to frustration. Yes the title is a bit misleading but we changed direction late in the game...
% - mention that although we collect facial expressions, these where done by iMotions and we would prefer to control derived data ourselves so we did not use it. 
% - What is our baseline random change? 50%? 33%? DONE 
% - make sure all tables and figures are referenced in the text! 
% - double check citation DOIs 
% - kill some darlings, murder some orphans DONE
% - paper is getting rather long - might need to cut down certain places! DONE 
% - tighten up the writing DOING 

Both academic and commercial visualization tools have invested heavily in user support over the years:
guided onboarding sequences~\cite{DBLP:journals/tvcg/DhanoaHFEGS25,stoiberPerspectivesVisualizationOnboarding2022}, tooltip-based explanations~\cite{DBLP:journals/ivs/ChunduryYCMSE23,stoiberDesignVisualizationOnboarding2023}, natural language query interfaces like Tableau's Ask Data, and recommendation engines that suggest chart types or data transformations.
However, a fundamental issue remains: how do we know \textit{when} to actually invoke them? 
Intervene too early---when the user is simply pausing to think, scanning the interface, or planning their next analytical move---and you get the Clippy effect: the infamous Microsoft Office paperclip assistant that interrupted users so aggressively it alienated an entire generation against proactive help systems~\cite{cooper_2007_about-face}.
Intervene too late, and the user has already lost focus, abandoned the task, or, worse, drawn incorrect conclusions from a visualization they misread but never questioned.
The fundamental challenge, then, is not \textit{what} help to offer, but \textit{when} to offer it.

In this paper, we propose the \textsc{Frustrometer}: a convolutional neural network (CNN) that predicts whether users are stuck or not during visualization tasks by fusing physiological and interaction signals. %into a single, continuous measure.
We ground the Frustrometer in a controlled study where $N=14$ participants performed 15 analytical tasks using two interactive visualization dashboards, at increasing difficulty. 
During each session, we captured a dense ensemble of physiological data, i.e., eye movement including fixations and saccades, pupil dilation, facial expression, head orientation, galvanic skin response (GSR), alongside interaction data: mouse position and velocity, keyboard events, and visualization-level interactions such as filtering, selection, and navigation~\cite{DBLP:journals/tvcg/YiKSJ07}.
We then segmented and annotated these sessions using both experimenter observation and participant self-reports to establish ground truth labels for frustration, stuckness, and level of user confidence.
The Frustrometer learns from this multimodal feature space to classify stuckness across multiple categories, with the best model achieving approximately 70\% classification accuracy ($F1$ score of 0.77). 
% We make no claim that 64\% represents deployment-ready accuracy.
With a binary classification problem where chance performance is 50\%, the Frustrometer performs meaningfully above baseline. 
% \future{TO BE VERIFIED!}
%% We validated our classifier through both standard precision-recall analysis on a held-out test\todoo{TODO} set and through detailed manual analysis using a custom temporal visualization tool we call \textsc{EyesOnData}, a web-based environment for aligning and exploring multiple synchronous data streams against segmented session recordings.
An ablation study focusing on individual data modalities shows that certain signals, primarily mouse and gaze dynamics, account for the majority of the classifier's predictive power across participants.
Other physiological signals contribute, but modestly. 
The multimodal configuration had superior accuracy, with unimodal mouse and gaze dynamics not far behind. 
This is the paper's most consequential finding: you may not need an eye tracker, a GSR sensor, and a blood pressure cuff to detect stuckness and frustration.
Tracking mouse movements may suffice, while adding gaze-tracking and a GSR sensor slightly improves accuracy. %  and possibly a webcam tracking facial expression may suffice.

We claim the following contributions in this paper:

\begin{enumerate}
    \item A \textbf{multimodal frustration dataset}\footnote{\tiny\href{https://osf.io/4vmc7/overview?view_only=9042007fc31f4ba99cc72342e107d20c}{https://osf.io/4vmc7/overview?view\_only=9042007fc31f4ba99cc72342e107d20c}} comprising physiological and interaction data from $N = 14$ participants performing 15 analytical tasks with two visualization dashboards (one with three common visualization types: barchart, stacked area chart, and scatterplot~\cite{ellemoseEyeBeholderMeasuring2026, lee_vlat_2017}, and one with two uncommon visualization types: node-link diagram and matrix-representation). 
    This data is annotated with observer and self-reported frustration labels and made publicly available.
    
%     \item A \textbf{feature framework} that maps raw sensor signals---eye tracking, GSR, pupil dilation, mouse dynamics, keyboard events, and visualization interactions---to frustration-relevant features suitable for classification.

    \item The \textbf{Frustrometer}, a CNN trained on multimodal data, which classifies user stuckness with approximately 70\% accuracy. 
    % A unimodal CNN 1D and CNN 2D both reach 73\% agreement from mouse movements alone, while unimodal CNN 2D and LSTM reach roughly 68\% agreement from gaze data alone.  

    \item An \textbf{ablation study} demonstrating that interaction patterns, particularly longitudinal gaze and mouse dynamics, carry the majority of predictive signal. Notably, in a unimodal configuration, mouse dynamics achieved the highest agreement, suggesting that lightweight instrumentation may be sufficient for real-time frustration detection in visualization systems.

\end{enumerate}

\section{Related Work}
The idea of detecting and addressing when the user feels frustrated, stuck, or experience other difficulties is not new%. 
%Here we review relevant literature on frustration and task difficulties, as well as different techniques and systems that has been previously proposed to address these issues. 
%% What is frustration, and why it is still relevant to talk about today [INTRO]
%Frustration arising from using a computer system is an old phenomenon
, however, it is still very relevant today \cite{hertzumFrustrationStillCommon2023}. 
% Definitions
Frustration can be defined as \textit{``irritable distress after a wish collided with an unyielding reality''} \cite[p.1]{jeronimusFrustration2017}. 
In other words, frustration arises when a desired outcome cannot be met because the circumstances do not allow it. 
In HCI, the desired outcome is typically the completion of some task, and the circumstances that allow or hinder this is the computer system in use. 
% Being stuck and not knowing how to proceed is a clear example of a circumstance that will likely lead to frustration. 
In such a case we might say that the user is stuck. 
`Being stuck' typically means that the user is unable to reach their goal, and is often followed by negative emotional states, such as frustration, as well as high mental demands~\cite{dreyBeNotBe2021}. 
However, `being stuck' is not black and white, but rather a spectrum, ranging from comfortably stuck, through moderately stuck, to truly stuck. 
At comfortably stuck levels the user might consider the task positively challenging, but when truly stuck, they feel frustrated and might abandon the task entirely. 

 % stuck is a spectrum --> this helps with our results in that the model detects stuckness at a certain level of the spectrum, but the level that is clearly visible to the onlooker might be more severe. 

%% Signposting
% In the remainder of this section we review where stuckness and frustration in visualizations come from, the existing techniques explore to mitigate it, ways of estimating the user's mental state, and how related fields address stuckness and frustrations.

%% Where does frustration come from in vis tasks? [SENSEMAKING PROCESSES]
\subsection{Sensemaking and Analytical Processes}
    %% \future{Coherent story and writeup}

    In visualizations we would expect that stuckness and the resulting frustration arise during the various sensemaking and analytical processes~\cite{DBLP:journals/tvcg/YiKSJ07}. 
    For novices, an unfamiliar visualization might result in the novice viewer \textit{floundering}, failing to create a meaningful frame for understanding and using the visualization~\cite{lee_how_2016}. Lee et al~\cite{lee_how_2016} only examined individual interactive visualizations in isolation, rather than multiple coordinated visualizations (MCVs) often used for sensemaking activities. Also, they did not ask the participants to solve concrete tasks, but rather try to understand how the visualizations encode the data. 
    However, it is not unlikely that the same mechanisms are at play for sensemaking activities, where multiple charts are involved, and more or less concrete tasks are performed. % MCVs with various cross-filtering options. 
    % Likewise, when tasked with solving concrete tasks, the user needs not only understand how the visualizations encode the data, but also how they might use the visualizations to find the answer. 

    The complexity of the visualizations~\cite{ellemoseEyeBeholderMeasuring2026, chuWhatMakesVisualization2026} used might also affect how easily the user ends up floundering. 
    Stuckness and frustration might arise from the visualization being too complex for the user, where again they might not be able to construct a frame for how to use a particularly complex visualization. 
    % resulting in the user floundering. 
    When solving tasks, complex visualizations, even if they are more effective for solving the task, could potentially hinder more than benefit the user. 
    Prior work suggest that the cognitive load of visualizations differ, not only between visualizations, but also between tasks they are used to solve \cite{hullmanBenefittingInfoVisVisual2011, saketTaskBasedEffectivenessBasic2019}. 
    In other words, the right tool is needed for the job.

    % In general, increase in task difficulty is linked to worse performance, and higher mental demands %% These thionigs are obvious and we dont have the space for it...
    % The effects of task difficulty and multitasking on performance \cite{adlerEffectsTaskDifficulty2015}
    % Effects of task difficulty and time-on-task on mental workload \cite{hagaEffectsTaskDifficulty2002}

%% How have we addressed potential frustration and stuckness already? [HELP & GUIDANCE]
\subsection{Help and Guidance in Visualization}
    %% \future{Coherent story and writeup} 

    One of the ways researchers have tried to address the potential floundering before it arise is through guidance~\cite{ceneda_characterizing_2017}, including onboarding~\cite{stoiber_visualization_2019}, and on-demand help systems~\cite{choeEnhancingDataLiteracy2024}. 

    In visual analytics, guidance has been described as a mixed-initiative system, where the user and the system provide feedback to each other in order to steer the analysis towards a common goal~\cite{cenedaReviewGuidanceApproaches2019}, effectively narrowing the knowledge gap that might keep the user from achieving their goal~\cite{ceneda_characterizing_2017}. 
    
    Guidance can play a number of roles~\cite{collins_guidance_2018}. 
    Onboarding users to new visualizations have been extensively explored~\cite{stoiber_visualization_2019, stoiberDesignVisualizationOnboarding2023, stoiber_abstract_2022, stoiber_comparative_2022, stoiber_design_2023}. 
    With onboarding, the goal is the make the user able to use the visualizations before they start their analytics tasks. 
    Different approaches have been explored, including guided tours~\cite{DBLP:journals/tvcg/DhanoaHFEGS25, hoque25dashguide}, contextual help~\cite{DBLP:journals/ivs/ChunduryYCMSE23, choe_enhancing_2024}, and visualization literacy scaffolding~\cite{stoiber_comparative_2022, choe_enhancing_2024}. 

    Advanced approaches to real-time, on-demand guidance has also been explored. 
    System such as Eviza~\cite{DBLP:conf/uist/SetlurBTGC16} and BOLT~\cite{srinivasan2023bolt} provides natural language support to users, while recent advancements in large language models allow these to be integrated into visualization systems, such as LEVA~\cite{zhao24leva}, LightVA~\cite{DBLP:journals/tvcg/ZhaoWXZGTZC25}, and ProactiveVA~\cite{zhaoProactiveVAProactiveVisual2026}. 

    While some guidance systems require explicit activation, others build a model of the user, in order to determine when to provide guidance. 
    Often the system will use sensory data to model the users emotional or cognitive state, e.g. to detect frustration~\cite{panwarProvidingContextualAssistance2018, cenedaShowMeYour2022}, 
    or use e.g. gaze and interaction data to determine areas of interest~\cite{srinivasanAttentionAwareVisualizationTracking2025, gadhavePredictingIntentSelections2021}. 
    
    % AAV proposes that if the viz is intelligent, then it respond to the needs of the user, 
    % How to provide assistance to a user of a visualization system has been studied. 
    
    % Ceneda et al.'s characterization of guidance in visual analytics~\cite{ceneda_characterizing_2017} and Collins et al.'s framework~\cite{collins_guidance_2018}
    
    % Onboarding approaches: guided tours~\cite{DBLP:journals/tvcg/DhanoaHFEGS25, hoque25dashguide}, contextual help~\cite{DBLP:journals/ivs/ChunduryYCMSE23}, visualization literacy scaffolding~\cite{stoiber_comparative_2022, choe_enhancing_2024} % generally a lot of work by stoiber is on onboarding approaches that should be referenced 

%% How might we detect that the user is stuck? [BIOMETRICS & INTERACTIONS]
\subsection{Detecting User State}
%\future{Coherent story and writeup}
%\USTP{If you know some examples of using e.g. ML to model the users state (mental, cognitive load, etc.) feel free to add them here.}

A large body of work has explored how to infer the emotional state of users from observable signals. These signals can be categorized into 1) \textit{external body signals} (e.g., gaze-based features, facial expressions, and body gestures), 2) \textit{internal body signals} (e.g., heart rate, galvanic skin response, blood pressure, and electroencephalography (EEG)), and 3) \textit{contextual signals} (e.g., interaction data such as speech or voice, as well as mouse and keyboard dynamics)~\cite{yang2021review}.

Among external body signals, gaze-based approaches are the most extensively investigated. Broader reviews link eye-tracking features such as fixations, saccades, and pupil dilation to attention, cognitive workload, and emotional processing, making them suitable for estimating cognitive and affective states~\cite{skaramagkas2021review}. Prior work has used such signals to predict confusion and emotion states in visualization settings~\cite{lalle2016predicting, sims2020neural, conati2020comparing, steichenInferringVisualizationTask2014}. Eye-tracking features have also been used to estimate task difficulty in data visualization~\cite{ferdousUsingPupilSize, liPredictingSpatialVisualization2020} and more generally to support the assessment of cognitive function~\cite{vulpe-grigorasiCAVIRCognitiveAssessment2025}. These findings suggest that moments of uncertainty or confusion during analysis often become visible in the user’s visual behavior before they are explicitly reported.

Internal body signals such as heart rate and GSR are commonly used as indicators for emotions and other affective states~\cite{saganowski2022emotion}. Taylor et al.~\cite{taylorUsingPhysiologicalSensors2015} used wearable physiological sensors to detect user frustration induced by variable system-response delays. In contrast to interaction data, physiological measures may capture more direct manifestations of frustration. At the same time, they typically require additional instrumentation and can be more sensitive individual variability and environmental conditions.

Among contextual signals, interaction-based approaches exploit data already available in most systems, including mouse dynamics, keystrokes, and touchscreen behavior. Reviews in affective computing show that such data can support emotion recognition without requiring specialized hardware~\cite{yang2021review}. Safaei et al.~\cite{safaeiSpeedDistanceExpanding2022} showed that cursor-based interaction features beyond simple speed and location measures, such as hesitation patterns, interaction pauses, click dynamics, and trajectory irregularities, can improve frustration detection from user behavior. In other domains such as education, interaction data has been used to detect when people learning a programming language are stuck~\cite{oka2023system}. For visualization systems, this is especially appealing because interaction logging is inexpensive, although the resulting signals could be ambiguous when interpreted in isolation.

Several studies combine multiple data modalities to leverage the benefits of different signal types. Lall\'e et al.~\cite{lalle2016predicting} utilized eye-tracking and interaction data to predict confusion in information visualization tasks. Panwar and Collins~\cite{panwar2018detecting} used eye-tracking and physiological measures to identify negative emotions induced by visual analytics tasks and, in a follow-up study, proposed a system for frustration detection to provide contextual assistance in visual analytics~\cite{panwar2018providing}. More generally, multimodal data has also been explored for predicting cognitive abilities of users for adapting interactive visualizations~\cite{conati2020comparing}. 

Overall, prior work shows that user state can be inferred from both unimodal and multimodal signals, typically utilizing supervised machine learning methods. Most studies have relied on traditional classifiers such as random forests and Bayesian classifiers~\cite{lalle2016predicting, panwar2018detecting, panwar2018providing, conati2020comparing, oka2023system}, while more recent work has begun to investigate deep learning approaches~\cite{sims2020neural}. Despite this progress, surveys of user-adaptive visualization continue to identify reliable user modeling as an open problem~\cite{yanez2025state}.

\section{Data collection}

    In order to gather a suitable dataset of physiological readings and interaction logs while users perform visual analytics tasks, we implemented a visualization tool and collected data from 14 participants performing 15 tasks each. 
    % Signposting
    In this section, we outline the implemented system, the tasks participants performed, and the protocol for data collection and subsequent processing.

\subsection{Apparatus}

    We implemented two interactive dashboards using Apache's ECharts.\footnote{https://echarts.apache.org/}
    Both dashboards visualized data about  top-rated Hollywood movies.\footnote{https://www.kaggle.com/datasets/rajugc/imdb-top-250-movies-dataset}
    We chose this topic because we expected that most people have some familiarity with movies, and at the same time this kind of metadata is not a topic that is likely to trigger an emotional response that would affect autonomous physiological functions such as heart-rate or sweat production. 
    
    The first dashboard consists of four common data visualizations \cite{ellemoseEyeBeholderMeasuring2026, lee_vlat_2017}: 
    (1) A barchart tallying the number of movies per genre; 
    (2) A stacked area chart showing the temporal distribution of movie releases per genre; 
    (3) A barchart showing the average box office performance per genre; 
    (4) A scatterplot of the IMDB rating vs. box office. 
    
\begin{figure*}[tbh]
        \centering
        \includegraphics[width=\linewidth]{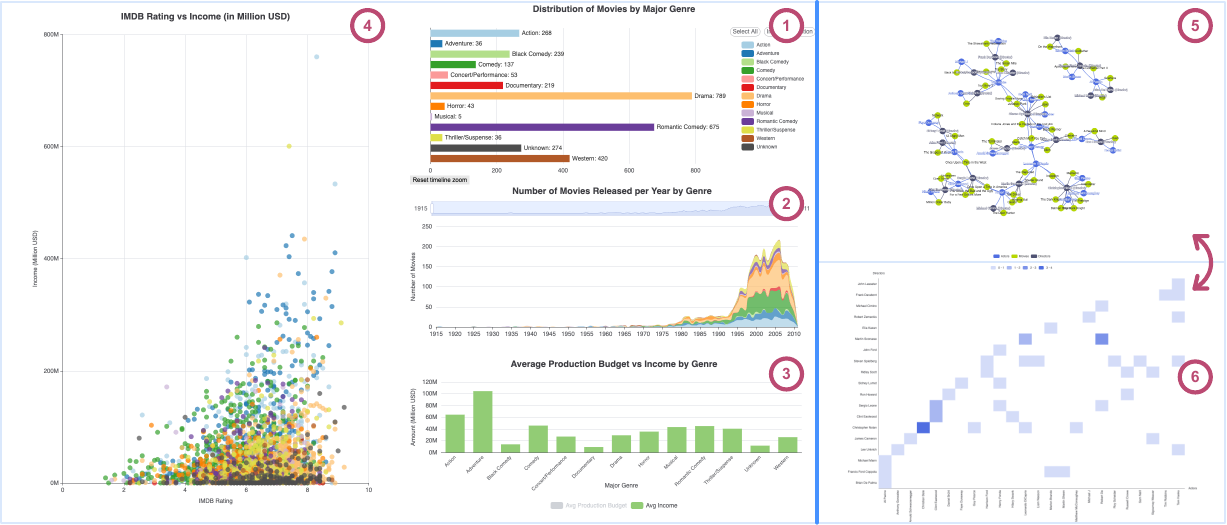}
        \caption{\textbf{Visualizations used for the task.} Participants completed nine tasks with the interactive dashboard (left), and six questions with the node-link and matrix visualizations (right). Questions were displayed in a sidebar to the right of the visualization(s) (omitted here).}
        \label{fig:dashboards}
    \end{figure*}

    The second dashboard consists of two uncommon data visualizations, (5) a node-link diagram showing connections between directors, lead-actors, and movies. Using a button, users can switch to (6) a matrix-representation of the director and lead-actor connections. 
    Figure~\ref{fig:dashboards} shows the visualizations used for the tasks.

%% This one ugly
    % \begin{figure*}
    %     \centering
    %     \begin{subfigure}[b]{0.68\textwidth}
    %         \includegraphics[width=\linewidth]{figs/dashboard1_clean.png}
    %         \caption{Interactive dashboard of movie data.}
    %         \label{fig:dashboard_1}
    %     \end{subfigure}
    %     \begin{subfigure}[b]{0.31\textwidth}
    %         \begin{subfigure}{\textwidth}
    %             \includegraphics[width=\linewidth]{figs/dashboard2_nodelink_clean.png}
    %             % \caption{Interactive Node-link and matrix representation.}
    %             \label{fig:dashboard_2_NL}
    %         \end{subfigure}
    %         \begin{subfigure}{\textwidth}
    %             \includegraphics[width=\linewidth]{figs/dashboard2_matrix_clean.png}
    %             % \caption{Interactive Node-link and matrix representation.}
    %             \label{fig:dashboard_2_M}
    %         \end{subfigure}
    %         \caption{Interactive Node-link and matrix representations of the same data.}
    %     \end{subfigure}
    %     \caption{\textbf{Visualizations used for the task.} Participants completed nine tasks with the interactive dashboard, and six questions with the node-link and matrix visualizations. Questions were displayed in a sidebar to the right of the visualization(s) (omitted here).}
    %     \label{fig:dashboards}
    % \end{figure*}

%% This one fire

    Interactions consisted of filtering and cross-filtering of categories, and hovering on any data-point for details. 
    In the first dashboard we used ECharts' \textit{datazoom} feature to allow the participants to filter movies by release year in the stacked area chart (2).
    In the second dashboard we added a button to change between the node-link diagram and matrix representation of the same data.

% task design 
\subsection{Task design}
    
    Tasks were designed to progressively increase in difficulty for the two different dashboards. 
    We operationalized task difficulty as the number of information cues the user needs to identify and process \cite{luoUserChoiceInteractive2019}. 
    Simple tasks were made up of mostly acquisition sub-tasks, such as identifying the longest bar in a barchart, while more difficult tasks could also require the user to evaluate multiple data points \cite{speierInfluenceInformationPresentation2006}. 
    Since our visualizations were interactive, we also added an increased need for interactions for the more difficult tasks, such as hovering for details, filtering, and cross-filtering. 
    The easiest task required no interaction. 
    Subsequent tasks were easier to solve with interactions, while the most difficult tasks required extensive interactions to be completed.

    All tasks were multiple choice with four possible answers. 
    A skip button allowed the participants an exit in case they could not find the answer to the task. 
    An overview of the tasks, as well as the identified optimal way to solve them is available in the supplemental materials. 
    
\begin{figure}
        \centering
        \includegraphics{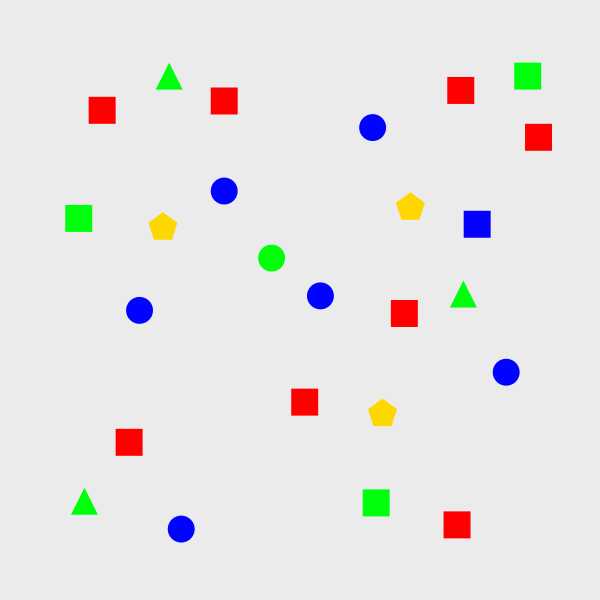}
        \caption{\textbf{Baseline figure used to gather baseline metrics.} Baseline tasks were performed using a simple image of a number of colored geometric shapes to simulate easy variants of lookup tasks typical in visual data  analytics tasks. Questions were displayed in a sidebar to the right of the image (omitted here).}
        \label{fig:baseline-task}
    \end{figure}

    A baseline assessment consisting of six tasks, designed to capture the physiological state of the participants prior to performing the tasks, serving as reference values for the physiological signals. 
    These tasks were designed to mimic the task the participants would encounter later but were kept very simple so as not to cause frustration.  The participants were asked to e.g. count the number of green elements on screen and identify the shape of the yellow elements from a picture with different colored shapes (see figure~\ref{fig:baseline-task}).

\subsection{Equipment and Data Collection}
\label{sec:equipment}

    % \future{Image of the data collection setting????? --- no space}
    
    The tasks were performed seated at 24-inch desktop computer monitor with a resolution of 1920 by 1080 pixels, at a normal viewing distance, with a standard wired computer mouse. 
    A Tobii Pro Nano eye tracker mounted under the display was used to collect gaze and pupil data. 
    An HD webcam was used to record a video of the participant’s face in order to collect data on facial expressions and head posture.  
    A heart rate sensor (Polar Belt H10) attached to the chest was used to record heart rate data (ECG), and a Shimmer3 EDA sensor (Galvanic Skin Response, GSR) attached to the wrist was used to record GSR data such as skin resistance and skin conductance. Both sensors were connected via Bluetooth to the computer used for the study.
    An overview of the equipment used is also shown in table~\ref{tab:ML_data_sources}. 
    
    Physiological data was captured via the iMotions software platform. 
    Interactions with the dashboards were captured by the system and saved as JSON files. 
    Mouse data was captured both by iMotions and the dashboards. 
    Dashboard data and iMotion data were synchrononised to iMotion's relative reference time, which is specified in milliseconds since the start of the experiment.

\subsection{Participants}
    
    We recruited 14 participants (50\% female) from a local study program, aged 20 to 29 (M=23, SD=3). 
    All participants had normal or corrected-to-normal vision, and none had any color blindness. One participant used glasses but only for distance viewing and not at the computer. 
    
    The participants were in an international class introducing them to data visualizations, which fit the target group of people who are not complete beginners but also far from experts of visualizations. 
    Three of the participants reported having experience with visual data analysis tools, 9 reported having some experience, and 2 reported being novices. 
    One of the participants reported having some experience with eye-tracking technology, while the rest reporting having little to no experience. 
    Participants where not compensated monetarily for their participation in the study.

\subsection{Protocol}
    
    Participants were greeted and introduced to the overall purpose of the study. 
    They were then asked to read and sign the consent forms, including optionally consenting to having the gathered physiological and interaction data be published as supplemental material. 
    Participants then filled out a short demographics form and were administered the MiniVLAT~\cite{DBLP:journals/cgf/PandeyO23} test on paper. 
    
    Next, the participants put on the heart-rate monitor and GSR sensors, before watching a short video explaining the different interactive elements they would encounter doing the task. 
    Finally, the structure of the study was explained in detail, including the self-assessment questions interleaved between the tasks.

    The participants explored the dashboard before starting the actual tasks. 
    This allowed participants to familiarize themselves with the interface, constructing a frame of the data.  
    Users pressed `next' to proceed to the tasks when they were ready. 

    Participants completed the tasks at their own pace. 
    After 1--1.5 minutes, the participant was asked if they wanted a hint on how to solve the task. 
    These hints were predefined. 
    Participants could ask for the hint at any time in case they initially rejected getting a hint. 
    Participants were asked to refrain from guessing in case they could not find the answer using the visualization, but rather skip the question. 
    
    After each task, participants were asked to rate their own confidence in their process on a seven-point Likert scale. 
    After three tasks, the participants were asked to rate these tasks using the NASA TLX on seven-point Likert scales. 
    This was because three tasks shared the same overall task difficulty estimate. 
    Participants completed nine tasks for the first dashboard. 
    For the second dashboard, due to the tasks not increasing noticeably in difficulty, but instead being uncommon visualizations, the NASA TLX was administered after all six tasks were completed for this dashboard.

    Participants completed the tasks, including exploration and self-assessments described above, for each dashboard. 
    At the end of the study, participants were thanked for their time, and offered some sweets. 
    The study took 30--53 minutes (Avg=44, SD=6.5), 
    with the tasks, including baselines, taking 15--32 minutes to complete (Avg=23, SD=5). 

\subsection{Data Processing}

    The data was processed in a number of ways before it was used for statistical tests and machine learning. 
    First, all data was synchronized to use iMotion's relative time format of milliseconds since the start of the session. 
    Additional preprocessing performed for the machine learning is described in sec~\ref{sec:ml_preprocessing}. 
    We report on our statistical findings in section~\ref{sec:statistics}. 
    We report on our machine learning experiments in section~\ref{sec:machine_learning}.

    \subsubsection{Observer Annotations} 

        % We wanted to have an observer look at the participants while they completed the tasks, and determine if and when they seemed stuck. 
        % This is inspired by activities such as pair-programming, where a second coder observes the user, and provides timely suggestions and feedback. 

        Inspired by activities such as pair-programming, where a second coder observes and provides timely suggestions and feedback, we had an observer look through the recordings and data streams of the participants while they completed the tasks, and mark if and when they seemed stuck. 
        In order to have an operational way of estimate when the participants seemed stuck we opted for a ternary definition, rather than a continuum. 
        Using this definition, the annotations were made using a coding guide with the following states (the full coding guide is available in the supplemental material): 
        \begin{description}
        \item
        [On-track] is described as `the person is executing actions that make clear progress toward completing the task.' Since we knew the task and the actions needed to solve them the participants were considered on-track when they performed these. Whenever they performed a series of actions confidently, even if it might not directly lead them towards solving the goal, we also considered this on-track, since the participants were not exploring solutions.
        
        \item
        [Uncertain/exploring] is described as `the person is trying things but without a clear direction.' We grouped being uncertain with exploring, since we found it difficult to differentiate the two, and since when exploring how to solve a task participants are uncertain---otherwise they would be on-track. % ---and vice versa. 
        
        \item
        [Stuck] is described as `When progress has stalled. Things like repeated failed attempts at the same action, long idle periods, erratic or seemingly random actions, or going in circles.' Importantly there seems to be no evidence of systematic exploration, but rather random attempts at solving the task. 
        \end{description}

        One author looked through the screen and webcam video feeds for all participants, and marked when the participant seemed uncertain or exploring how to solve the task, when they were on-track towards completing the task, and when they seemed stuck. 
        The annotations resulted in the participants being stuck 13\% of the task time, while the remaining time was roughly split between being uncertain/exploring or on-track.

        %% IRR 
        In order to validate these annotations, we performed a Time-Segmented Cohen's Kappa Inter Rater Reliability analysis for six tasks from one participant with a researcher not involved with the project. 
        The segment size was set to 1 second. 
        Agreement between the two Annotators was 66.2\%. 
        Annotators often agreed on the state of the participant, but the specific time-stamp that the change from uncertain to on-track or stuck, and vice versa was never agreed upon exactly. 
        The Cohen's Kappa coefficient of 0.431 results in a moderate agreement between annotators. 
        While the results are not perfect, we felt confident that our annotations reflect the overall state of the participants' with regards to being stuck or on-track. 
        
        % Percent agreement:  66.2%  (141657/214111)
        %   Cohen's Kappa:      0.431
        %   Interpretation:     Moderate agreement

%%%%%%%%%%%%%%%%%%%%%%

    \subsubsection{Summary} 

        From the data collection study and the data processing, we have the following data: 

        \begin{enumerate}
            \item \textbf{Physiological and interaction measurements.} These are the data streams captured during the data collection study, and include gaze, GSR, ECG, and mouse interaction data. 

            \item \textbf{Participant self-assessments.} These include the confidence the participant had per task, as well as the NASA TLX questionnaire collected for every three tasks. 

            \item \textbf{Expert annotations of the participants' stuckness.} These are the labels of different segments of the tasks, labeled as either `on-track', `uncertain/exploring', or `stuck'. 

            \item \textbf{Task difficulty.}  These are the estimates of the task's difficulty. Tasks 1--3 are low difficulty, tasks 4--6 are medium difficulty, and tasks 7--9 are high difficulty. Tasks 10--15 have a medium difficulty, but use an uncommon representation. 
        \end{enumerate}

\section{Statistical Analysis}
\label{sec:statistics}

We performed a series of statistical tests to examine correlations between self-assessments and observer annotations. 
% With this analysis we wanted to investigate whether higher frustration/lower certainty correlate with what we see as external observers, such as the percentage of time being stuck or on-track. 
Strong correlations mean that observer annotations can be a proxy for the participants' self-assessments of e.g. frustration and performance, meaning self-assessments are not needed to model these. 
We expected that there is some correlation between these metrics, as spending more time being stuck would likely be reported as requiring more effort, higher mental demand, higher levels of frustration, and lower confidence and performance. 
We did not expect physical demand, or temporal pressure to be affected, as the tasks required the same amount of physical labor, and the tasks had no time pressure. 

We also examined correlations between task difficulty and self-assessments, and between task difficulty and observer annotations. 
If task difficulty is apparent from self-assessment data and from observer annotations, we acn estimate the difficulty of the task, another potential source of stuckness and frustration. 
We expected that more difficult tasks result in more time being stuck, with worse self-reported performance, and more frustration, higher mental demand and effort required to complete the task. 
We again did not expect physical demand, or temporal pressure to be affected, as the tasks required the same amount of physical labor, and the tasks had no time pressure.

\subsection{Comparing Self-assessments to Expert Annotations}

\begin{figure}
        \centering
        \begin{subfigure}[b]{0.32\linewidth}
            \includegraphics[width=\linewidth]{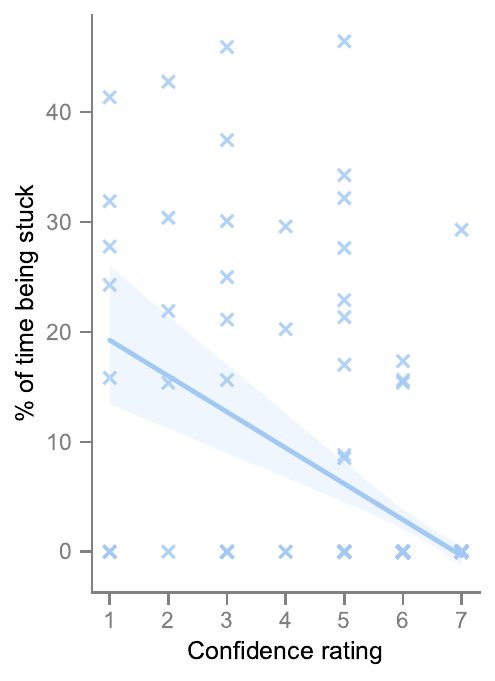}
            % \caption{Confidence rating plotted against the \% of time being stuck.}
            % \label{fig:rating-vs-stuck}
        \end{subfigure}
        \begin{subfigure}[b]{0.32\linewidth}
            \includegraphics[width=\linewidth]{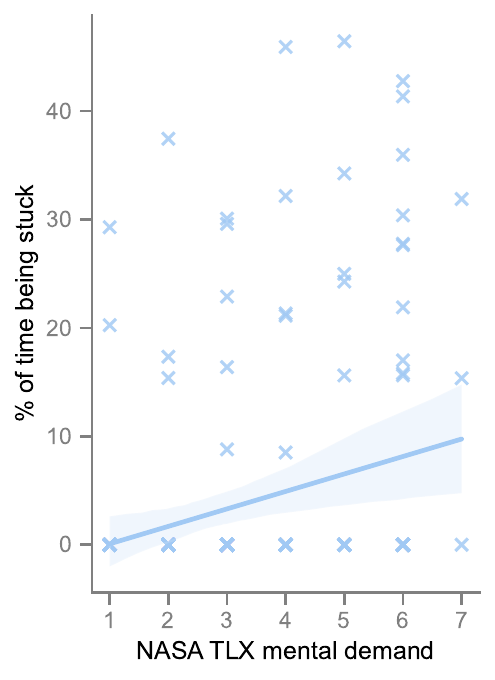}
            % \caption{Mental demand plotted against the \% of time being stuck.}
            % \label{fig:mental-vs-stuck}
        \end{subfigure}
        \begin{subfigure}[b]{0.32\linewidth}
            \includegraphics[width=\linewidth]{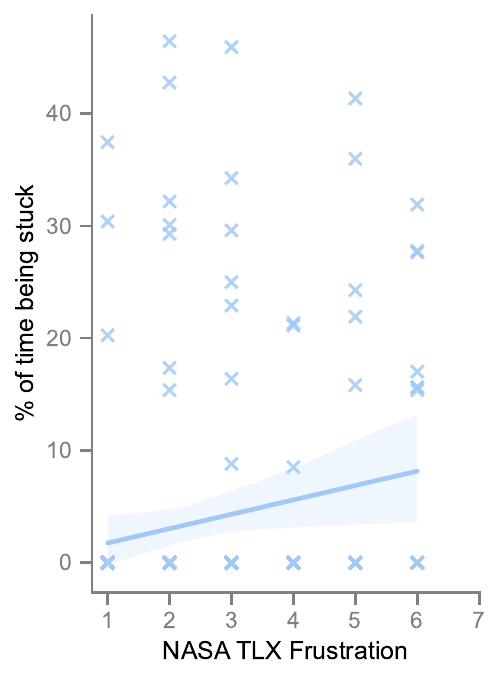}
            % \caption{Frustration level plotted against the \% of time being stuck.}
            % \label{fig:frustration-vs-stuck}
        \end{subfigure}        
        \caption{\textbf{Self-assessments vs \% of time being stuck.} 
        These were the only results form the within-subject Spearman test that had significant correlations. 
        Confidence rating has a much higher (negative) correlation than mental demand and frustration, which have similar correlations with the annotations. 
        }
        \label{fig:stuckness-stats}
    \end{figure}

    We performed a within-subject Spearman (one-sample Wilcoxon) with Benjamini-Hochberg FDR correction, between the participant self-assessments, and the percentage of time spend per task being stuck, on-track, and uncertain respectively. 
    Results revealed only three statistically significant correlations with the percentage of time being stuck: Confidence rating, mental demand, and frustration (See figure~\ref{fig:stuckness-stats}.

    Confidence rating was significantly negatively affected by longer periods of being stuck (median $r_s = -.48$), with a one-sample Wilcoxon signed-rank test confirming that correlations were significantly greater than zero, ($W = 0.0, p_{fdr} = .010, n = 12$). 

    Mental demand was significantly positively affected by longer periods of being stuck (median $r_s = .29$), with a one-sample Wilcoxon signed-rank test confirming that correlations were significantly greater than zero, ($W = 2.0, p_{fdr} = .041, n = 12$).

    Frustration level was significantly positively affected by longer periods of being stuck (median $r_s = .32$), with a one-sample Wilcoxon signed-rank test confirming that correlations were significantly greater than zero, ($W = 2.0, p_{fdr} = .041, n = 12$).

    All results can be found in Appendix~\ref{tab:self_assessment_vs_annotations_spearman}. These results indicate that being stuck affects the confidence in the participants' ability to complete the task, as well as increase their mental demand and frustrations. 
    This is somewhat expected, as the being stuck in a task would likely lead to the user feeling frustrated, their confidence dropping, and their mental demand increase. 

    We had expected to also the inverse effect; that more time being on-track would result in less frustration, less mental demand, and more confidence, however the data does not support this. 
    % Since there is relatively little time being spend stuck (13\% of task time)

\subsection{Comparing Task difficulty to Self-assessments \& Expert Annotations}

    We performed Friedman tests between the percentages of time participants were stuck, on-track, and uncertain per tasks versus the different self-assessments and expert annotations respectively, 
    followed by a post-hoc pairwise Wilcoxon signed-rank test with Bonferroni-correction of the significant pairs. 
    All results can be found in Appendix~\ref{tab:difficulty_vs_annotations_friedman}. 
    % , 
    % \ref{tab:difficulty_vs_assessments_friedman},
    % \ref{tab:difficulty_vs_self-assessments_wilcoxon}, and
    % \ref{tab:difficulty_vs_annotations_wilcoxon}.

\begin{figure*}
        \centering
        \begin{subfigure}[b]{0.138\linewidth}
            \includegraphics[width=\linewidth]{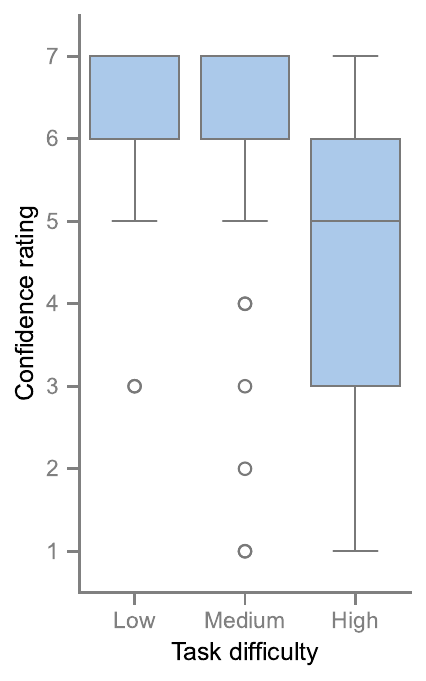}
            % \caption{Interactive dashboard of movie data.}
            % \label{fig:rating-vs-stuck}
        \end{subfigure}
        \begin{subfigure}[b]{0.138\linewidth}
            \includegraphics[width=\linewidth]{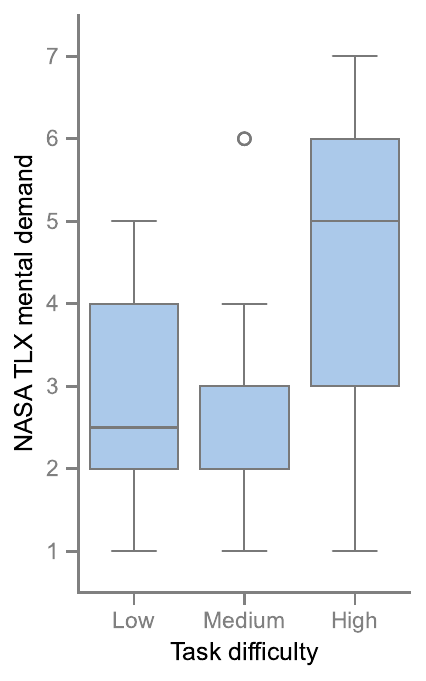}
            % \caption{Interactive dashboard of movie data.}
            % \label{fig:rating-vs-stuck}
        \end{subfigure}
        \begin{subfigure}[b]{0.138\linewidth}
            \includegraphics[width=\linewidth]{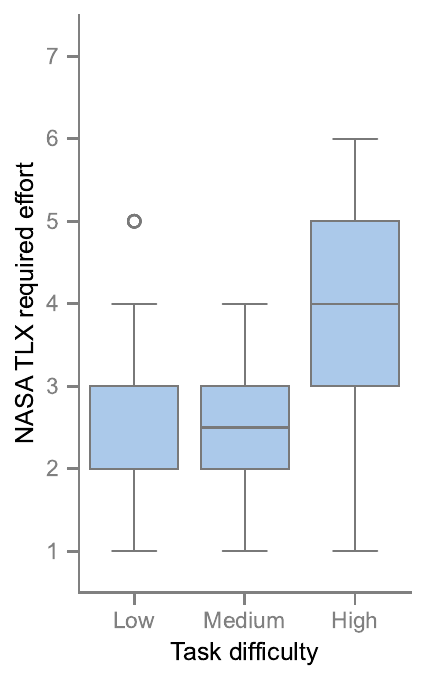}
            % \caption{Interactive dashboard of movie data.}
            % \label{fig:mental-vs-stuck}
        \end{subfigure}
        \begin{subfigure}[b]{0.138\linewidth}
            \includegraphics[width=\linewidth]{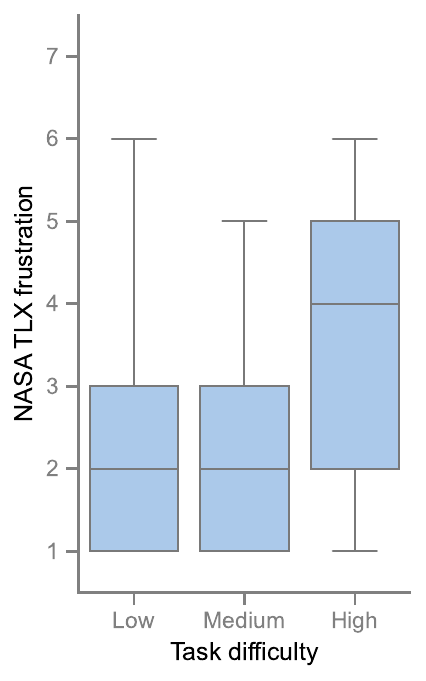}
            % \caption{Interactive dashboard of movie data.}
            % \label{fig:frustration-vs-stuck}
        \end{subfigure}        
        \begin{subfigure}[b]{0.138\linewidth}
            \includegraphics[width=\linewidth]{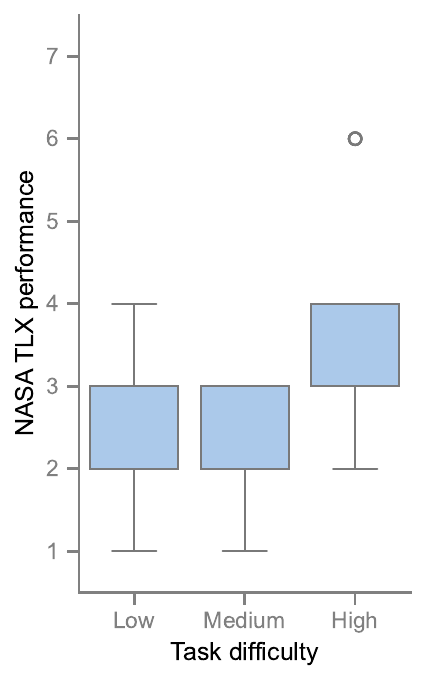}
            % \caption{Interactive dashboard of movie data.}
            % \label{fig:rating-vs-stuck}
        \end{subfigure}
        \begin{subfigure}[b]{0.138\linewidth}
            \includegraphics[width=\linewidth]{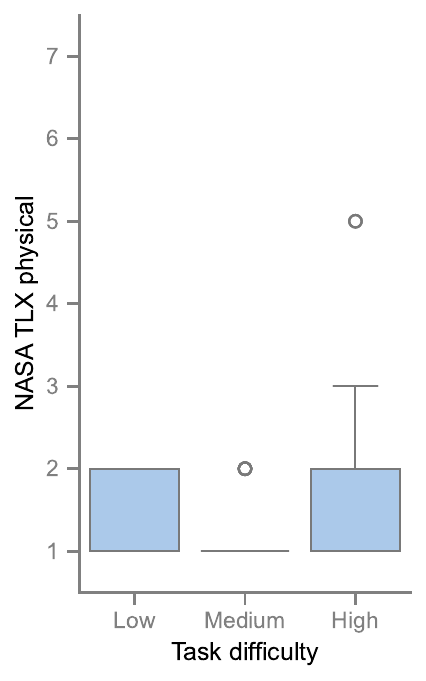}
            % \caption{Interactive dashboard of movie data.}
            % \label{fig:rating-vs-stuck}
        \end{subfigure}
        \begin{subfigure}[b]{0.138\linewidth}
            \includegraphics[width=\linewidth]{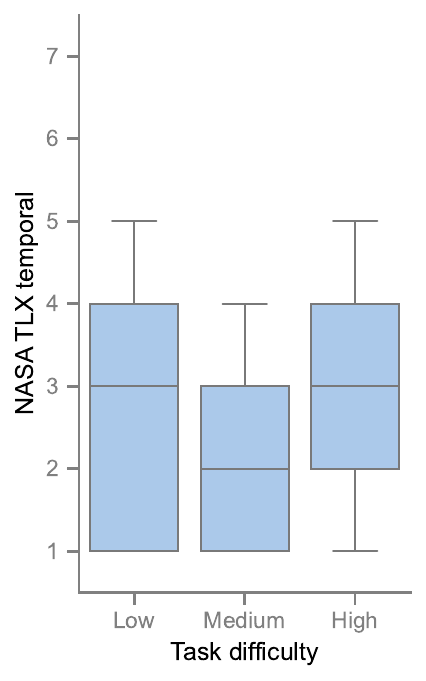}
            % \caption{Interactive dashboard of movie data.}
            % \label{fig:rating-vs-stuck}
        \end{subfigure}
        \caption{\textbf{Task difficulty vs self-assessments.} 
        % \niklas{Need to fill out the caption.}
        For confidence ratings, mental demand, required effort, frustration, and performance, the high difficulty tasks have significantly different self-assessments from the low and medium difficulty tasks respectively, but not between low and medium difficulty tasks. 
        Neither temporal demand nor physical demand were significantly different for the three task difficulties. 
        }
        \label{fig:difficulty-self-assessments-stats-individual}
    \end{figure*}

    Results show that some self-assessments metrics can be useful to determine the relative difficulty of the task, however the results are not conclusive (see Figure~\ref{fig:difficulty-self-assessments-stats-individual}). 
    Of the self-assessments, especially confidence, mental demand, and required effort were useful to distinguish between low and high difficulty tasks, as well as medium and high difficulty tasks. 
    None of the self-assessments showed significant difference between log and medium difficulty tasks. 

\begin{figure}
        \centering
        \begin{subfigure}[b]{0.32\linewidth}
            \includegraphics[width=\linewidth]{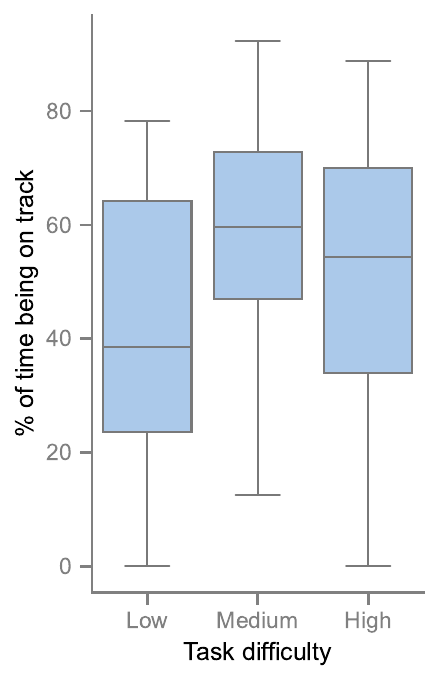}
            % \caption{Interactive dashboard of movie data.}
            % \label{fig:rating-vs-stuck}
        \end{subfigure}
        \begin{subfigure}[b]{0.32\linewidth}
            \includegraphics[width=\linewidth]{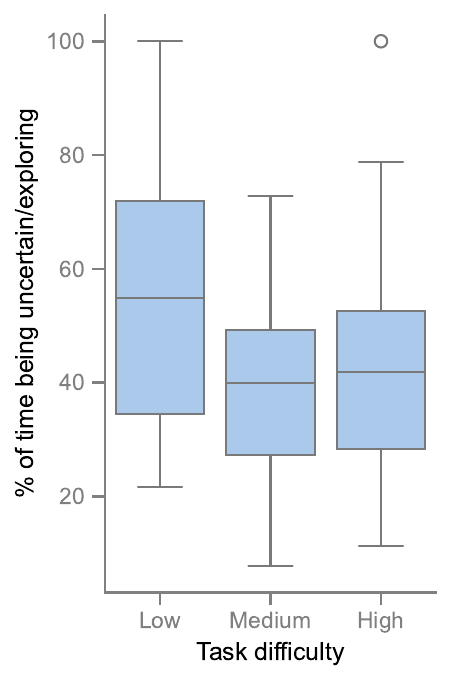}
            % \caption{Interactive dashboard of movie data.}
            % \label{fig:rating-vs-stuck}
        \end{subfigure}
        \begin{subfigure}[b]{0.32\linewidth}
            \includegraphics[width=\linewidth]{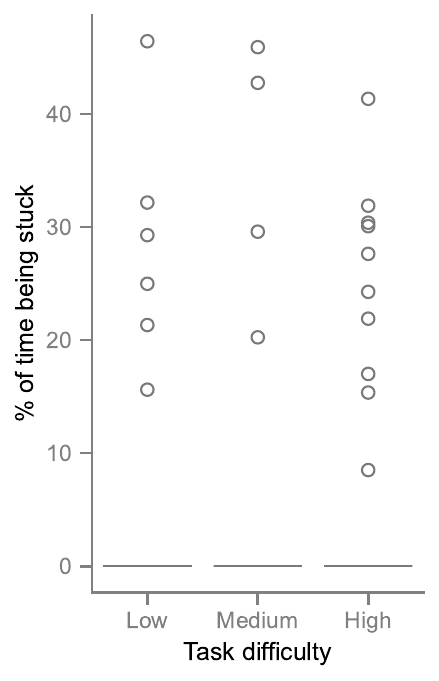}
            % \caption{Interactive dashboard of movie data.}
            % \label{fig:rating-vs-stuck}
        \end{subfigure}
        \caption{\textbf{Task difficulty vs expert annotations.} 
        % \niklas{And fill out this one too!}
        There was a significant relation between the \% of time being on-track for low and medium difficulty tasks, but not between high difficulty tasks and any of the other task difficulties. 
        There were significant relations between the \% of time being uncertain for the low difficulty tasks and the other two task dificulties. % It is higher for low difficulty, which might be a learning bias, since we do not have that much time of the participants being stuck that it is the whole story! 
        }
        \label{fig:difficulty-annotations-stats}
    \end{figure}

    Expert annotations showed significant differences in the amount of time being on-track and uncertain/exploring between low and medium difficulty tasks, however not for the other pairwise task difficulties (see Figure~\ref{fig:difficulty-annotations-stats}).

    % \begin{figure*}
    %     \centering
    %     \includegraphics[width=\linewidth]{figs/difficulty-vs-all.pdf}
    %     \caption{\textbf{Task difficulty vs self-assessments.}}
    %     \label{fig:difficulty-self-assessments-stats-most}
    % \end{figure*}

    % \begin{figure*}
    %     \centering
    %     \includegraphics[width=\linewidth]{figs/difficulty-vs-all-all.pdf}
    %     \caption{\textbf{Task difficulty vs self-assessments.}}
    %     \label{fig:difficulty-self-assessments-stats-all}
    % \end{figure*}

    % \subsection{Results}
    
    % \future{Here we should probably analyse the results a bit for what they mean for our ML model etc. \\
    % - How the results relate to the def of stuckness and frustration. \\
    % - What does it tell us about the ability to detect stuckness and frustration. 
    
    % --- no cause that is a discussion type of thing}

\section{Machine Learning Experiments}
\label{sec:machine_learning} 

We performed extensive and detailed machine learning experiments on the collected data, using the observer annotations as ground truths. 
Here we detail these experiments, including data preprocessing, model evaluation, and modality ablation studies.

\subsection{Data Preprocessing and Dataset Construction}
\label{sec:ml_preprocessing}

        % intro to data preprocessing - to avoid repeated headings?? 
    
       %\subsubsection{Preprocessing and Normalization}

        After synchronizing the raw multimodal sensor data, we normalized the data using within-subject $z$-score standardization to account for individual physiological differences. For each participant, the mean and standard deviation were computed from their baseline recording phases. Participant 1 was excluded from the dataset  due to missing heart rate data, ensuring a complete and consistent multimodal dataset.

    %\subsubsection{Dataset Construction}
        
        To capture transient cognitive states, we applied a sliding window approach with a 5-second window size and a 2.5-second step (50\% overlap). For the machine learning analysis, we focused on a binary classification setting distinguishing between ``stuck'' and ``on-track'' states. Each window was assigned a ground-truth label only if at least 50\% of its duration overlapped with the corresponding manual annotation, i.e., at least 2.5 seconds annotated as either ``stuck'' or ``on-track''. Participants 2, 6, and 13 were excluded from the dataset because they did not exhibit any episodes of ``stuck'' behavior, resulting in 10 participants in the final dataset.
    
        % \textbf{Stuck:} Windows containing at least 2.5 seconds (50\% of the window duration) of annotated ``stuck'' behavior.
        
        % \textbf{Not-Stuck:} Windows requiring at least a 50\% overlap with periods explicitly annotated as ``On-track.''
        
        To prevent the machine learning model from adopting a majority-class bias or overfitting to participants who spent disproportionately longer in a specific state, the dataset was balanced intra-subject. For each participant, we randomly under-sampled the majority class to exactly match the sample count of the minority class, ensuring equal representation of \textit{stuck} and \textit{on-track} states per user.
            
    %\subsubsection{Dataset Construction: Static and Temporal Formats}
        
        To support both traditional machine learning and deep learning architectures, the windowed data was processed into two distinct formats: 
        
        \paragraph{Tabular input data.} For traditional machine learning models we calculated a comprehensive set of statistical aggregations across the time-series data within each window. 
        The extracted features include the mean, standard deviation, minimum, and maximum for continuous physiological signals such as GSR and heart rate, as well as for eye-tracking data and head pose orientations. 
        Furthermore, discrete event frequencies were quantified by extracting fixation counts, saccade counts, and blink rates. 
        Heart rate variability was assessed directly from raw R-R intervals by calculating the Root Mean Square of Successive Differences (RMSSD) and the percentage of successive normal sinus intervals greater than 50 milliseconds (pNN50).
        Kinematic trends were captured by calculating linear slopes for continuous signals like GSR to indicate directional shifts. Mouse interactions were summarized by the total distance moved and the ratio of idle time within the given window.

        To ensure the model utilizes generalizable features rather than overfitting to noise or individual user traits, we applied a feature selection pipeline. Features dominated by inter-subject variance were removed to prevent the model from learning who the participant is rather than what state they are in. Any feature where the between-subject variance ratio exceeded 0.5 was excluded. This filtered out highly user-specific physiological ranges, including maximum, minimum, and mean heart rate, as well as all raw GSR aggregations. Furthermore, we computed the correlation matrix for all features and pruned pairs with highly redundant distributions ($|r| > 0.9$). For instance, fixation count ($r = 0.95$ with saccade count) was removed. Absolute positional features, such as head\_pitch\_mean, were discarded as they often reflect an individual's default posture rather than state-driven movement.

        The final feature set used to train traditional machine learning models includes 14 features demonstrating significant class separability (Cohen’s $d$ ranging from 0.25 to 0.49, $p < 0.001$). 
        The full results are available in Appendix~\ref{tab:ML_feature_selections}.

        \paragraph{Time series input data.} For the deep learning methods, we used the raw synchronized time-series data sampled at 60~Hz as input. Each sample was represented as a two-dimensional matrix comprising 300 time steps and multiple channels.
        These channels included continuous heart rate, GSR, eye-tracking and head-pose data, as well as mouse-velocity differentials along the x- and y-axes. 
        For certain deep learning architectures, we additionally computed Short-Time Fourier Transform (STFT) spectrograms directly from the temporal window data.
        
        %The final dataset formatted to allow sequential models to directly learn the temporal evolution and complex inter-modality interactions preceding a specific cognitive state (see table~\ref{tab:ML_feature_selections})
        %\USTP{Reference Table with feature selection ^^ is this reference correct??}

        \begin{table}
            \small
            \centering
            \begin{tabular}{llrl}\toprule
                 \textbf{Modality} & \textbf{Data Source} & \textbf{Sampling rate} & \textbf{Imputation method} \\\midrule
                 GSR &  Shimmer3 EDA &  123 Hz & Linear Interpolation\\
                 Eye Tracking &  Tobi Pro Nano &  60 Hz & -\\
                 Heart Rate &  Polar H10 BLE &  1.5 Hz & Linear Inerpolation\\
                 Mouse Dynamics &  Custom JS Logger &  ca 6Hz & Forward Fill\\ \bottomrule
            \end{tabular}
            \caption{\textbf{Data sources and imputation methods} used for the machine learning experiments. A description of the data collection equipment can be found in section~\ref{sec:equipment}.}
            \label{tab:ML_data_sources}
        \end{table}
        
\subsection{Classification Models and Evaluation Strategy}
    
    To evaluate the generalizability of the predictive models to unseen users, we employed a Leave-One-Subject-Out (LOSO) cross-validation strategy for all experiments. Model performance was evaluated iteratively on the held-out participant, and overall performance was quantified using the average classification accuracy across all test folds, as well as macro-averaged precision, recall, and $F1$ score.

    Prior to model training, the 14 filtered features were mean-centered per participant.
    Within each iteration of the LOSO cross-validation loop, a standard scaler was fit strictly on the training partitions and applied to the held-out participant to strictly prevent data leakage. Similarly, the time-series data were mean-centered per participant and standard scaling was applied independently to each sensor channel within the boundaries of the cross-validation loop.
    
    The evaluated traditional machine learning methods include Support Vector Machines (with an RBF kernel), $k$-Nearest Neighbors, Multi-Layer Perceptrons, Decision Trees, Random Forests, and Gradient Boosting. For each model, a hyperparameter grid search was conducted. 
    %The evaluated algorithms included Logistic Regression (L1 and L2 regularized), Support Vector Machines (RBF kernel), K-Nearest Neighbors, Multi-Layer Perceptrons, and various tree-based architectures ranging from single Decision Trees to advanced ensembles (Random Forests, AdaBoost, Gradient Boosting, HistGradientBoosting, XGBoost, LightGBM, and CatBoost). 
     
    The evaluated deep neural network architectures included  a Long Short-Term Memory (LSTM) network, a one-dimensional CNN, a hybrid CNN-LSTM architecture, and a two-dimensional CNN operating STFT spectrograms. For architectures including LSTM layers, the search covered hidden size, the number of LSTM layers, dropout, and learning rate; for CNN  architectures, it covered the number of filters, kernel size, dropout, and learning rate. All deep learning models were trained for a maximum of 100 epochs utilizing the Adam optimizer. 
    The loss was calculated using binary cross-entropy with logits, incorporating dynamic positive class weighting to dynamically adjust for any residual batch-level class imbalances. 
    To optimize convergence and prevent overfitting, gradient clipping (norm of 1.0) was enforced, and an early stopping mechanism was implemented to halt training if the validation $F1$ score failed to improve for 15 consecutive epochs.
    The model state yielding the highest validation $F1$ score during the training run was restored for the final evaluation on the held-out test participant.

    Following the hyperparameter grid searches, the optimal configurations for all 10 models (six traditional machine learning models and four deep learning models) were evaluated using LOSO cross-validation.

    \paragraph{Results.} The classification results for the mutlimodal setting are summerized in Table~\ref{tab:ML_ranks_full}. 
    Overall, the deep learning models outperformed the traditional machine learning baselines, particularly with respect to the $F1$ scores, indicating better balanced performance across classes. The best overall accuracy and $F1$ scores were achieved by the 2D~CNN operating on STFT spectrograms (69.7\% and 0.768, respectively), suggesting that time-frequency representations capture highly informative multimodal patterns. In contrast, the standard LSTM achieved the highest accuracy (69.8\%), indicating strong overall classification performance on the raw time-series input. Among the traditional models, Gradient Boosting performed best in terms of accuracy and $F1$ (68.8\% and 0.669, respectively), while the remaining approaches achieved generally lower $F1$ scores. 
    
    \begin{table}
        \small
        \centering
        \begin{tabular}{llcccc}\toprule
            \textbf{Model} & \textbf{Type} & \textbf{Acc.} & \textbf{Precision} & \textbf{Recall} & \textbf{$F1$}\\\midrule
            2D CNN& Spectrograms& 0.697 & 0.671 & 0.933 & 0.768\\
            CNN--LSTM          & Time series& 0.686 & 0.654 & 0.828 & 0.727\\
            LSTM               & Time series& 0.698 & 0.693 & 0.754 & 0.718\\
            Gradient Boosting  & Tabular & 0.688 & 0.707 & 0.650 & 0.669\\
            1D CNN             & Time series& 0.625 & 0.622 & 0.646 & 0.626\\
            KNN                & Tabular & 0.633 & 0.636 & 0.615 & 0.621\\
            SVM RBF            & Tabular & 0.636 & 0.644 & 0.592 & 0.612\\
            MLP                & Tabular & 0.586 & 0.552 & 0.713 & 0.610\\
            Decision Tree      & Tabular & 0.625 & 0.663 & 0.556 & 0.591\\
            Random Forest      & Tabular & 0.624 & 0.606 & 0.572 & 0.584\\\bottomrule
        \end{tabular}
        \caption{\textbf{Classification results for traditional machine learning and deep learning models.} Reported performance metrics include macro-averaged classification accuracy (Acc.), precision, recall, and $F1$ scores.}
        \label{tab:ML_ranks_full}
    \end{table}
    
    %\subsubsection{Final Evaluation and Inter-Subject Analysis}
        % Following the hyperparameter grid searches, the optimal configurations for all 10 models (six traditional machine learning models and four deep learning models) were locked for final evaluation. 
        % The models were evaluated using the established Leave-One-Participant-Out cross-validation pipeline. 
        % To provide a comprehensive assessment of predictive performance, we calculated the aggregate Accuracy, Micro F1, and Macro F1 scores (see Table~\ref{tab:ML_ranks}).
        
        % Furthermore, to investigate the generalizability of the learned features across diverse users, we conducted a inter-subject analysis. 
        % For each held-out participant, the prediction accuracy was recorded across all 17 optimized models. 
        % Analyzing the distribution of accuracies per participant allows for the identification of specific users whose cognitive or physiological expressions of the ``stuck'' state deviate significantly from the population norm, providing crucial insight into the overarching robustness of the multimodal feature space and the boundaries of user-independent modeling.
        % \USTP{do we have some table or figure etc of these results we can bedazzle the reviewers with?, Yes, will be insertet shortly}
    
    \subsection{Ablation Analysis of Data Modalities}
        To quantify the individual predictive power of each sensor modality (i.e., eye tracking, heart rate, GSR, head movement, and mouse kinematics), we conducted an ablation study. For the tabular input data, each modality was represented by its corresponding set of features; for GSR, this additionally included all features that had previously been removed during the feature selection pipeline. For the time-series input data, only the raw sensor channels corresponding to the targeted modality were used. All models (with hyperparameters determined during the aforementioned best-model selection) were trained and evaluated on the respective modality-specific data using LOSO cross-validation.

\begin{figure}
        \centering
        \includegraphics[width=\columnwidth]{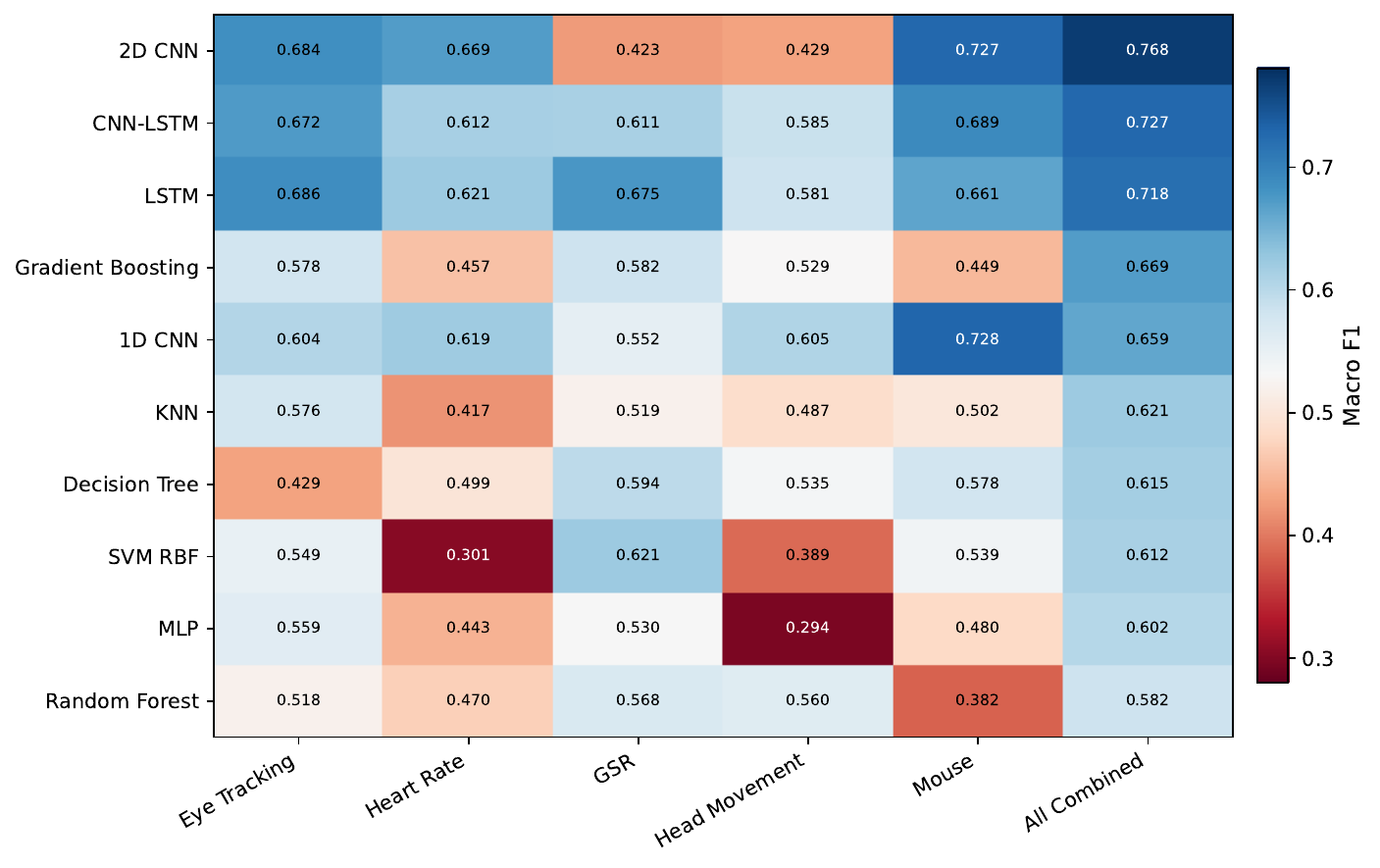}
        \caption{\textbf{Ablation study results} in terms of macro-averaged $F1$ scores, showing the predictive power of the individual unimodal settings. Results are ranked according to the $F1$ scores obtained in the multimodal setting.}
        \label{fig:ablation-study}
    \end{figure}

        \paragraph{Results.}
        The modality ablation results (see figure~\ref{fig:ablation-study}) show clear differences in predictive power across sensor types. The multimodal setting yielded the strongest overall performance, with an average $F1$ of 0.657 across all evaluated models, followed by eye tracking ($F1 = 0.585$), mouse kinematics ($F1 = 0.574$), and GSR ($F1 = 0.568$). In the multimodal setting, the best-performing model was the 2D~CNN on spectrograms ($F1 = 0.768$). Moreover, the multimodal configuration outperformed the best unimodal setting for most models (SVM with GSR and 1D~CNN with mouse kinematics being the only exceptions). Among the unimodal settings, mouse kinematics was the strongest overall modality, achieving its highest result with 1D~CNN ($F1 = 0.728$), followed closely by eye tracking, with its best result obtained by the LSTM ($F1 = 0.686$); GSR, also performing best with the LSTM ($F1 = 0.675$); and heart rate, which achieved its highest result with the 2D~CNN ($F1 = 0.669$). The weakest unimodal setting overall was head movement, whose best result was obtained with 1D~CNN ($F1 = 0.605$). Across most modality settings we can observe that deep learning models achieved higher overall performance than the traditional machine learning models.    
        Overall, these results indicate that multimodal fusion is generally beneficial, while eye tracking, mouse kinematics, and GSR provide the most informative unimodal signals.

       %Performance was aggregated using Accuracy, Macro F1, and Micro F1 scores to directly compare the efficacy of single-modality models against the fully integrated multimodal baseline. 

\section{Discussion}

In this section, we discuss our findings, and what they mean for detecting stuckness and the resulting frustration experiences by stuck users of visualization systems. 
First we explain our results in detail, then we evaluate the goal of this paper: whether stuckness, and the resulting frustration, can be detected using physiological and interaction signals, by a human onlooker, as well as through machine learning.
We elaborate on the challenges of detecting stuckness, before illustrating design implications from our findings. 
Finally we address the limitations of our work, and point out future research directions.

\subsection{Explaining the Results} %% Maybe this is too on the nose? 

% - what does the stats results mean 
%     - statistical correlation between rating and % stuck, and between metnal demand and frustration and the % stuck 
%           - we can use % stuck as a standin fo their self-assessment!! 
%     - more difficult tasks can be detected using self-assessments and more importantly from the annotations (which we can detect using ML, but taht is a secret for the next section) 
% - what does the ML results mean
%     - We can detect the difference between on-track and stuck 
%     - Uncertain/exploring is difficult. 
%     - some modalitise matter more than others 
%         - ablation study results 
%         - some modalities lower performance when combined compared to standalone. 
%             - why could that be

% STATISTICAL ANALYSES

    \paragraph{Statistical analysis.} 
    
    The purpose of the statistical tests was to identify whether there are significant correlations between the participants' self-assessments of their performance and an observer's annotations of the same. 
    Strong correlations mean that observer annotations can be used in place of participants’ self-assessment. 
    This means that we do not need training sessions where users provide self-assessments to fine-tune a model, but can directly estimate their stuckness and performance from the amount of time they are e.g. stuck, on-track, provided that these annotations can be accurately determined from the physiological and interaction data. 
%
    % These results are important for the further analysis in this paper, as we are interested in estimating when the user becomes stuck without having them tell us first through a training session. 
    %
    %
    % If we are able to mimic the results of an observer observer, and these observations correlate with the users' own feeling of stuckness and frustration, then we do not need the user to self-assess their performance. 
    % 
    In addition, we wanted to investigate whether there are statistically significant differences in self-assessments across tasks of different difficulty levels. 
    This would allow us to estimate task difficulty based on observation alone, providing another indicator that the user might be interested in assistance. 
%     If we can estimate whether a task seems difficult for the user, it is another indicator that the user might be interested in assistance at some point. 
    
    Our results show that there are some correlations between the obsever annotations and the participants' self-assessments, specifically between the percentage of time the observer annotator determined that the participants were stuck and their own assessment of confidence, mental demand, and frustration levels (see Appendix~\ref{tab:self_assessment_vs_annotations_spearman}).
    The results show a good correlation between the confidence rating and the percentage of stuckness, and a moderate correlation between frustration levels and the percentage of stuckness, and a moderate correlation between mental demand and the percentage of stuckness. 
    These results are not surprising, since being stuck leads to frustration, and would inversely affect participants’ confidence. 
    Similarly, being stuck in a task would likely result in higher mental demand than a task where the participant is on-track throughout the task. 
    % It also makes sense that more difficult tasks impose a greater mental demand, where tasks difficult enough to result in some participants being stuck would also have a higher mental demand than tasks that do not result in participants being stuck. 
    % This correlation is also moderate. 
    One reason for the stronger correlation between confidence and percentage of stuckness, is that the participants were asked to rate their own confidence after every task, while the TLX questionnaire was administered after every third task, since the tasks were grouped by difficulty. 
    While this was done to reduce questionnaire fatigue, it also means that the TLX answers are less timely compared to the confidence self-assessments. 
    % 
    % While the self-assessments were based on a classic TLX questionnaire, the annotations were instead focused on external signs of the participant being on-track, stuck, or somewhere in-between. 
    % For the purposes of offering timely assistance, it is important to know whether the participants also felt that they were struggling, since if they were not, they would likely not want assistance.  
    From this statistical analysis, we can conclude that it is possible, to some extent, to estimate participants' self-assessments from the annotations of an external observer.

    Our second statistical analysis reveals that there are significant differences between participants' self-assessments as well as the observer's annotations across different task difficulties (see Appendix~\ref{app:statistics}). 
    Interestingly, none of the self-assessments showed significant differences between low difficulty tasks and the other task difficulties, while confidence ratings, mental demand, required effort, frustration, and performance all differed significantly between low and high difficulty tasks, and medium and high difficulty tasks. 
    Neither temporal pressure nor physical demand showed significant differences across difficulty levels. 
    These results are not surprising, since the TLX measures task load. 
    Since all tasks required the same amount of physical effort and imposed no time limit, neither temporal demand nor physical demand was  expected to be high for any difficulty level. 
    
    For the observer annotations, the results are less encouraging. % good. 
    The percentage of time spent \textit{on-track} differs significantly between low and medium difficulty tasks. 
    Similarly, the percentage of time spent \textit{uncertain/exploring} differs significantly between low and medium difficulty tasks, as well as between low and high difficulty tasks. 
    The percentage of time spent \textit{stuck} does not differ significantly across task difficulties. 
    In other words, few pairwise relations between observer annotations and task difficult have significant differences
    These results are not encouraging with respect to estimating the task difficulty from the observer annotations. 
    % They show that task difficulty cannot determine task difficulty from an ML system that is able to detect whether the user is stuck. 

% MACHINE LEARNING ANALYSES
    \paragraph{Machine learning analysis.}
    The machine learning results show that user state can be predicted above the random baseline (accuracy of 50\%), particularly in the binary setting distinguishing \textit{on-track} from \textit{stuck} behavior. Across the evaluated models, the deep learning methods generally outperformed the traditional machine learning methods, with the strongest results obtained by the 2D~CNN operating on STFT spectrograms, followed by the CNN--LSTM and the LSTM. This suggests that the temporal structure of the raw signals, and in particular their time-frequency characteristics, contain predictive information that is not fully captured by handcrafted features. However, the reasonably good performance of some traditional models suggests that the target state is also reflected in the relatively simple features obtained from external and internal body signals as well as contextual signals. Overall, our findings show that multimodal deep learning approaches, particularly those leveraging temporal or spectral structure, are better suited for the task at hand than traditional machine learning classifiers based on handcrafted features.
    
    The ablation study further shows which data modalities contribute most strongly to the prediction of stuckness. Multimodal fusion yielded the best overall performance, indicating that the different modalities provide complementary information. Among the unimodal settings, eye tracking, mouse kinematics, and GSR were the most informative, whereas heart rate and head movement contributed substantially less when used in isolation. This results suggest that stuckness is primarily expressed through observable interaction and attention behavior, while physiological arousal may provide supporting affective information. In particular, the consistently strong performance of models trained on eye-tracking and mouse kinematics data aligns with the idea that difficulties during analytical reasoning become visible in how users inspect the interface and how fluently they interact with it. These results also align with prior studies showing that eye-tracking and interaction signals are informative for identifying user difficulties and confusions as well as affective states. By contrast, the weak contribution of heart rate suggests that this signal may be too unspecific, too noisy, or too strongly influenced by inter-individual variability to serve as a robust indicator of stuckness in our setting.

    A further important observation is that the machine learning models are able to distinguish the ends of the stuckness spectrum; however, they struggle to classify its intermediate phases. In the binary setting, the models can detect relatively clear \textit{on-track} and \textit{stuck} states, whereas introducing the more ambiguous class \textit{uncertain/exploring} substantially reduces performance to only $F1 = 0.43$ (achieved by the CNN--LSTM and Random Forest), leading to many confusions with the existing two classes. For more detailed results of the multiclass classification, refer to the Appendix (Table~\ref{tab:3_class_ML_ranks}). This suggests that exploratory or transitional phases share characteristics with both \textit{on-track} and \textit{stuck} classes, making them harder to separate using the available signals. 

      \begin{figure}
        \centering
        \includegraphics[width=\columnwidth]{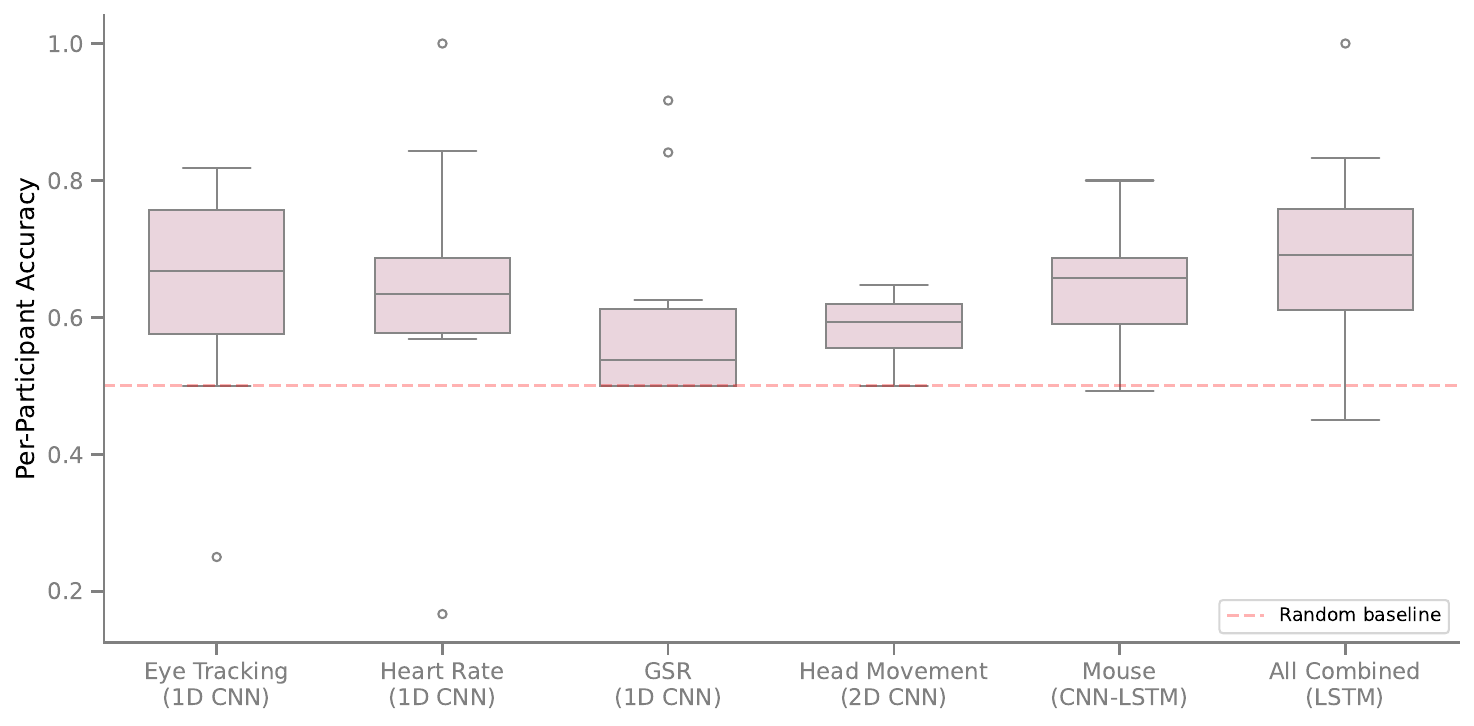}
        \caption{\textbf{Per-participant accuracy across individual modalities.} Accuracies were obtained using leave-one-subject-out cross-validation and are shown for each modality and the best-performing machine learning model. The dashed line denotes the random baseline.}
        \label{fig:ablation-study-boxplot}
    \end{figure}
    
    Lastly, Figure~\ref{fig:ablation-study-boxplot} highlights substantial inter-participant variation across modalities, indicating that stuckness and frustration are not expressed uniformly and that no single modality consistently achieves strong performance for every user. However, the high predictive power of eye-tracking data, and especially of the multimodal approach, suggests that we may be able to detect emerging difficulties before they become obvious to a human observer, opening up possibilities for earlier  intervention.

    % From the machine learning experiments...

    % We can detect when the user in either on-track or stuck---the ends of the stuckness spectrum---but the middle parts of the spectrum are too difficult to differentiate from either of these more clear classes, and including this third class results in much worse performance, where the model miss-classifies both on-track and stuck segments as uncertain/exploring. 

    % Stress indicators seem to be individual, as there is no signal that is the most important across all participants. 
    % This means that there is a need for more adaptive approach, that learns from the user's signals and which ones that are the most important. 
    
    % Gaze fixation was consistently among the most important features. 
    % This is inline with out expectations; That gaze is a very relevant feature to detect when the user is stuck. 
    
    % Heart-rate was never among the important features, which suggests that heart-rate is not needed to detect user stuckness. 
    
    % we might be able to detect the frustration before it even is visible to the onlooker. 

\subsection{Detecting when the User is Stuck }
% ==> Main discussion: Can we detect stuckness? 
% - qualified yes, under certain circumstances 
% Do we need a ground truth? -- Probably yes, for the calibration of biomarkers a baseline is useful 

% Modalities matter
% - strengths of a multimodal approach - robustness and redundancy 

% Minor points: 
% - Some models are really good at detecting some participants, but fail at others. 
%     - we are all different, so an adaptive system is likely the best option if deployed. 
%         - different modalities are useful for some participants and not for others 

    It is time to address the main goal of this paper: 
    \textit{Can we detect when the user is stuck or frustrated while working with data visualizations?}
    The short answer is yes. 
    The machine learning results show that we are able to distinguish between states in which the user is \textit{stuck} and states in which the user is \textit{on-track} with relatively high accuracy.  
    Meanwhile, we were unsuccessful in distinguishing the rather ambiguous \textit{uncertain/exploring} state without detrimentally affecting the accuracy of the other two classes. 
    This is not necessarily a limitation in practice. 
    For an intelligent guidance system, the primarily objective is to detect when a user has crossed from acceptable difficulty into  stuckness, so that assistance can be provided before frustration leads them to abandon the task. 
    For this purpose, the exact level of stuckness is less relevant than recognizing when intervention becomes necessary. 
    % However, this level is likely not the same for every person. 
    
    An important finding is that stuckness may be detectable through physiological signals before users are even consciously aware of it. This opens up the possibility of offering guidance not only when users feel stuck, but even before the negative effects of frustration set in. 

    We have chosen to structure our experiments to mimic a human observer's ability to determine whether and when a person is getting stuck while using visualizations, rather than rely purely on the user's own judgment of frustration and confidence. 
    As our statistical findings show, there is a correlation between participants' self-reporting and the observer's annotations, meaning that detecting when the user gets stuck % reaches an impasse 
    may also provide an indication of when they are experiencing some level of frustration. 
    % Main discussion point: 
    %     Is it possible to detect/estimate stuckness via ML like a human observer can do? 
    %     - biometrics + interactions 
    
    %     qualified yes under our specific restrictions (stuck vs on-track, rather than three classes). 
    %     - even under real-time 
    %     - on-track vs stuck - but not the in-between where they are exploring or are uncertain. 
        
    %     do we need a ground truth? -- probably for calibration we need some baseline for the biomarkers 

    % Sub discussion points 
    
    %     Earlier onset of biometric and interaction indicators for the ML model than what the human observer can see --> is it possible that we can actually detect frustration when it is incoming, and what does this mean for future proactive and adaptive support systems 

    %     Games as a perspective on different approaches to completing visualization tasks. 
    %     - speed-running vs exploring 

    % Explain the miss-classifications of the model...

\subsection{Implications for Design} %% Two implications, to keep it nice, short and tidy. 
% Design implications 
% - the mouse might be all you need! 
%     - no need for fancy equipment to detect frustration at a relatively good level

    % \future{Now that we've done it, what can we use it for?}

    % [How this signal can inform the design of adaptive guidance systems for complex visualization tasks]%% From the abstract & introduction

    Our findings show that everyone expresses frustration slightly differently. 
    For some users, frustration is visible in their facial expression, whereas for others it manifest more in their gaze or mouse movement behavior. 
    A well-designed guidance system would need to adapt to the individual, perhaps by learning which signals are most relevant for a given user and prioritizing those.  
    
    Similarly, people adopt different strategies when completing a visual analysis task. 
    Some participants carefully consider the available options before interacting with the visualizations, whereas others would quickly start interacting with the visualizations, looking for a way to solve the tasks. 
    In tasks that required the identification of a relevant data point,
    % before using hover to get details needed to solve the tasks, 
    some participants would let go of the mouse and visually inspect the data visualizations on the screen, while  
    % returning to use the mouse when they had found a data point of interest. 
    others would rapidly hover over multiple data points looking for clues in the data that could help them solve the task. 
    An adaptive approach should take these different strategies into account, for example by de-prioritizing mouse kinematics for the former group, and prioritizing it for the latter. 

    Furthermore, our multimodal approach provides redundancy and higher robustness. 
    If eye tracking is not available, the system’s accuracy may decrease, but its functionality will not be lost entirely. % not lose it completely like a system that relies purely on gaze data. 
    The multimodal approach allows the accuracy of the assistance to scale with the available input modalities.
    At a basic level, the system could rely only on mouse and interaction tracking. With access to a standard webcam, it could additionally incorporate facial expressions. If eye tracking is also available, the system could produce even more accurate estimates. 
    It is plausible to envision that data from, e.g., smartwatches and possibly EEG-enabled earphones could also be added to the array of input signals, providing many different streams of data for estimating the user’s state. 
    This approach also gives agency to the user, who can choose which data to provide to the system, thereby trading estimation accuracy against privacy and control.

    % Ethical considerations??? 
    %     Potential downsides: 
    %         we are using it for good, but just as we can detect frustration/stuckness it is not far of to see this being use to detect frustration or enjoyment, and tailor experiences to keep the user engaged for a longs as possible curating content based on their real-time estimated mental state. 

\subsection{Limitations and Future Work}
% Limitations: 
% - longer tasks with more realistic stakes 
% - more participants 
% - more vis and more diverse tasks 
%     - realistic setting with overarching goal where tasks build on each other? 
% - more modalities - EEG, facial expression etc. 
    
    % \future{What are some of the limitations and how will we address them in the future?}

    In the following, we address some of the limitations of our work, and point out directions for future research into adaptive guidance systems for visualizations.

    % Adding more modalities. 
    % EEG. 
    % Facial expression could be used - we chose not to since we did not calculate them ourselves and because those findings are highly related to the accuracy of the machine learning model used to extract facial expressions. 
    
    % Our study has a limited number of participants, with tasks only taking around 30 minutes on average to complete. 
    % Longer sessions with more participants would provide more data for training. 
    % Not all our participants got stuck. 
    % One way to force people to get stuck would be to introduce unsolvable tasks or hidden mechanisms, in order to induce stuckness and frustration. 
    % However such approaches might not reflect real-world situations. 
    
    % , not long-tern usage
    
    Introducing longer, more diverse tasks, and potential incentives for good performance and correctness may be one way to increase participants’ investment, make frustration due to being stuck more apparent, and generate additional training data. 
    % Unsolvable tasks
        % - pros and cons of such an approach
    We chose not to do this in this paper because we wanted to see frustration arise from a non-stressful situation, where external factors would not influence how much stuckness and frustration the participants experience. 
    However, external pressure to perform could mimic real-world decision-making situations, where people are required to both be fast and precise, and where getting timely assistance can be especially impactful.

    % different personalities and approaches to how to do visual analytics. 

    Given that the ground truth annotations are based on video recordings, the label boundaries are inherently imprecise. 
    Our IRR showed that while annotators tended to agree on the label, the exact timestamp for the label differed. 
    This means that some of the false positives of the machine learning models may reflect changes in the sensor data that are not yet externally visible to a human observer. 
    Employing an ensemble of annotators and deriving ground truth from their combined annotations could result in more robust labels, but this would be a time-consuming and laborious process. 
    
    %% Merge the above and below point somehow to make it shorter?? 

    %% Relatively important point, should be kept 
    Critically, we do not know whether the participants would agree that the annotations where correct or whether they actually needed help in the segments labeled as \textit{stuck}. 
    The only instances where we know for certain that they needed help are those in which they explicitly asked for a hint. 
    % Even the cases where they were offered the hint and accepted we cannot say for certain that they needed the hint, since they might just accept it to make the task easier for them, without it being truly needed. 
    % Some participants did reject the hint initially, and then either solved the task on their own or asked for the hint later, and only when they asked for the hint can we with certainty say that the participant needed assistance. 
    A study in which hints are provided more contextually, either by an automated system (which requires some of the findings from this and related papers) or by a human, could help clarify the circumstances under which people want assistance assistance and how this can be modeled even more precisely. 
    Although such a study would be time-consuming, it would offer great insight into when and why people want assistance. 
    Similarly, reviewing the stuckness labels from an automated system with participants could help fine-tune the model.

%% Move to discussion?? 
    Even then, outcomes are unlikely to be uniform across individuals, as some are willing to endure and find the solution on their own, where others will seek out assistance more quickly~\cite{luoHowUsersPerceive2026}. 
    The thresholds at which stuckness becomes critical~\cite{dreyBeNotBe2021} are different across people. 
    The situation in which assistance is offered is also very important. 
    Under time pressure, users might prefer more assistance to complete the tasks faster. In more leisurely settings, or when users are deeply invested in solving things themselves, such as in a video game, they may prefer less assistance and instead value the sense of achievement that comes from solving difficult tasks on their own. % REF?? 
    Further investigation into when people want assistance, and how to adapt to user desires could help making adaptive assistance systems achieve widespread adoption.

    Finally, the performance of our models depends partly on the preprocessing, which includes normalizing the data to the baseline tasks per participant. 
    While this was necessary to yield greater accuracy, requiring a baseline phase to finetune a classifier would not be practical in real-world settings.
    Instead, the system could gradually learn the behavioral patterns of the user over time.

    % group of annotators for robustness of the stuckness annotations 

    % participants were not asked directly after each task how frustrated are you, nor were they asked during the tasks if they were on-track, felt stuck etc. An experience sampling study could reveal this, but it might add to the felt frustration of being constantly interrupted. 
    % This is a fundamental compromise that researchers have to make when collecting emotional data from participants. 

    % %% Minor points we do not have space for. Parts of it can be put into the above point.
    % how to decouple the participants reactions to the underlying data from their interactions with the data visualizations as a tool. 
    % - we chose movie data because we consider it relatively safe from strong emotional and thus autonomous physiological responses. 
    % When the task data can lead to emotional responses, the accuracy of stuckness- and frustration-detecting models might degrade. 
    % Investigating how emotions affect model performance is relevant to increase robustness. 

\section{Conclusion} %  and Future Work 

In this paper we explore different techniques to detect and estimate stuckness in user working with data visualizations. 
We conducted a controlled study to collected multimodal physiological and interaction data with 14 participants, performing increasingly difficult tasks with interactive visualizations. 
Additionally the participants self-assessed their own performance throughout the study. 
This data was supplemented by observer annotations of the perceived stuckness of the participants. 

We then performed a series of statistical tests to determine the correlations between self-assessments and observer annotations, and between task difficulty, and self-assessments \& observer annotations respective. 
These tests reveal that annotations correlate with self-assessments, meaning that external observations can be used to estimate users' self-assessed performance. 

We then performed extensive machine learning experiments, using the observer annotations as the ground truth of stuckness. 
CNN approaches outperformed traditional ML approaches, with a  multimodal configuration yielding the most accurate results on average at roughtly 70\% accuracy, with gaze tracking, mouse kinematics and GSR data being the most performative single modalities across the examined models. 
Notably, the best performing single modality, 1D~CNN mouse kinematics, is only slightly behind the top performing multimodal configuration (2D~CNN). 
This suggests that the mouse kinematics alone might suffice to detect stuckness at reasonable accuracy. 
On the same note, our data indicate that heart-rate and head movement data is not particularly useful for detecting stuckness in users. 

In conclusion we are successfully able to detect stuckness in user, which is a precursor to frustration. 
Our results indicate that we might even detect stuckness before it becomes apparent for the user. 
Based on our results, we propose that a multimodal guidance system allows for more robust and adaptive guidance systems, that can adopt the available sensors to allow for the most accurate estimations. 
At the same time, our data shows that people demonstrate stuckness and frustration in different ways, meaning that truly intelligent guidance systems should be able to learn from, and adapt to the individual user's signals. 

%% Some rounding off statement here 
%  The ability to accurately detect when bla bla bla

\section*{Acknowledgements} % Interim acknowledgements for the review version

% \paragraph*{Usage of Generative AI}
% The code used for the expert annotations tool and for statistical analyses was partly written by Anthropic's Claude Opus 4.6 in extended thinking mode. 

\acknowledgments{% % Only for the camera-ready version. 
    This work was partly supported by Villum Investigator grant VL-54492 by Villum Fonden, and by the Austrian Research Promotion Agency (FFG) as part of the EyeQTrack project (COIN Aufbau FO999898083). % USTP grants 
    Any opinions, findings, and conclusions expressed in this material are those of the authors and do not necessarily reflect the views of the funding agencies.

    \paragraph*{Usage of Generative AI} 
    The code used for the expert annotations tool and for statistical analyses was partly written by Anthropic's Claude Opus 4.6 in extended thinking mode. 
}

\FloatBarrier

%% ---------------------------------------------------------------------
%% References
%% ---------------------------------------------------------------------
\bibliographystyle{abbrv-doi-hyperref} 
\bibliography{frustrometer}

\clearpage
\appendix 
\section{Appendix}

\subsection{Statistical experiments supplemental results}
\label{app:statistics}

Results form the statistical tests performed. Statistically significant differences and correlations are highlighted.

\subsubsection{Self-assessments vs expert annotations results}

\textbf{Results from the within-subject Spearman test} between self-assessments and expert annotations. Statistically significant correlations are highlighted.\\

\noindent
% Within subject spearman test between self-assessments and expert annotations
%\begin{table*}[]
% \centering
% \caption{\textbf{Results from the within-subject Spearman test} between self-assessments and expert annotations. Statistically significant correlations are highlighted.}
\label{tab:self_assessment_vs_annotations_spearman}
% \resizebox{\columnwidth}{!}{%
%\rotatebox{90}{%
\begin{tabular}{@{}llrrcrcc@{}}
\toprule
\textbf{Self-assessment} & \textbf{Annotation} & \textbf{Mean r} & \textbf{Median r} & \textbf{n (participants)} & \textbf{W} & \textbf{p (raw)} & \textbf{p (FDR)} \\ \midrule
\rowcolor[HTML]{DAE8FC} 
Confidence rating & \% stuck & -0.525 & -0.483 & 12 & 0.0 & 0.0005 & 0.0103 \\
\rowcolor[HTML]{DAE8FC} 
NASA\_TLX\_mental & \% stuck & 0.272 & 0.293 & 12 & 2.0 & 0.0058 & 0.0409 \\
\rowcolor[HTML]{DAE8FC} 
NASA\_TLX\_frustration & \% stuck & 0.268 & 0.322 & 12 & 2.0 & 0.0058 & 0.0409 \\
NASA\_TLX\_temporal & \% stuck & 0.256 & 0.231 & 10 & 2.0 & 0.0152 & 0.0598 \\
NASA\_TLX\_frustration & \% on-track & -0.236 & -0.242 & 13 & 12.0 & 0.0171 & 0.0598 \\
NASA\_TLX\_effort & \% stuck & 0.246 & 0.299 & 12 & 8.0 & 0.0122 & 0.0598 \\
Confidence rating & \% on-track & 0.251 & 0.296 & 14 & 19.0 & 0.0353 & 0.1058 \\
NASA\_TLX\_temporal & \% on-track & -0.236 & -0.357 & 12 & 13.0 & 0.0425 & 0.1115 \\
NASA\_TLX\_performance & \% stuck & 0.169 & 0.244 & 12 & 14.5 & 0.0640 & 0.1493 \\
NASA\_TLX\_frustration & \% uncertain & 0.142 & 0.155 & 13 & 19.5 & 0.0803 & 0.1687 \\
NASA\_TLX\_effort & \% on-track & -0.172 & -0.305 & 14 & 27.0 & 0.1189 & 0.2270 \\
NASA\_TLX\_temporal & \% uncertain & 0.139 & 0.189 & 12 & 20.5 & 0.1763 & 0.3085 \\
NASA\_TLX\_effort & \% uncertain & 0.066 & 0.132 & 14 & 31.0 & 0.1937 & 0.3129 \\
NASA\_TLX\_mental & \% on-track & -0.085 & -0.080 & 14 & 32.0 & 0.2166 & 0.3248 \\
NASA\_TLX\_performance & \% on-track & -0.087 & -0.154 & 13 & 25.0 & 0.2719 & 0.3807 \\
NASA\_TLX\_physical & \% stuck & 0.119 & 0.110 & 6 & 5.0 & 0.3125 & 0.4102 \\
NASA\_TLX\_performance & \% uncertain & 0.015 & 0.116 & 13 & 33.0 & 0.6374 & 0.7873 \\
NASA\_TLX\_physical & \% uncertain & -0.111 & -0.031 & 6 & 8.0 & 0.6875 & 0.8021 \\
NASA\_TLX\_mental & \% uncertain & -0.023 & -0.023 & 14 & 49.0 & 0.8552 & 0.8980 \\
Confidence rating & \% uncertain & -0.049 & -0.099 & 14 & 49.0 & 0.8552 & 0.8980 \\
NASA\_TLX\_physical & \% on-track & 0.025 & -0.015 & 6 & 10.0 & 1.0000 & 1.0000 \\ \bottomrule
\end{tabular}%
% }
% }
% \end{table*}

\vspace{2cm}

% \FloatBarrier
\subsubsection{Task difficulty vs self-assessments \& expert annotations results}

\textbf{Results from the Friedman test between task difficulty and self-assessments.} Statistically significant differences are highlighted.\\

\noindent
% Results from the Friedman test between task difficulty and self-assessments
% \begin{table}[]
% \centering
% \caption{\textbf{Results from the Friedman test between task difficulty and self-assessments.} Statistically significant differences are highlighted.}
\label{tab:difficulty_vs_assessments_friedman}
\begin{tabular}{@{}lrcr@{}}
\toprule
\textbf{Self-assessment} & \textbf{$\chi^2(2)$} & \textbf{\textit{p}} & \textbf{Kendall's \textit{W}} \\ \midrule
\rowcolor[HTML]{DAE8FC} 
Confidence rating & 11.472 & 0.0032 & 0.294 \\
\rowcolor[HTML]{DAE8FC} 
NASA\_TLX\_mental & 6.980 & 0.0305 & 0.179 \\
\rowcolor[HTML]{DAE8FC} 
NASA\_TLX\_effort & 16.930 & 0.0002 & 0.434 \\
\rowcolor[HTML]{DAE8FC} 
NASA\_TLX\_frustration & 7.610 & 0.0223 & 0.195 \\
\rowcolor[HTML]{DAE8FC} 
NASA\_TLX\_performance & 8.750 & 0.0126 & 0.224 \\
NASA\_TLX\_temporal & 1.771 & 0.4124 & 0.045 \\
\rowcolor[HTML]{DAE8FC} 
NASA\_TLX\_physical & 6.125 & 0.0468 & 0.157 \\ \bottomrule
\end{tabular}
% \end{table}

\vspace{2cm}

\noindent
\textbf{Results from the Friedman test between task difficulty and expert annotations.} Statistically significant differences are highlighted.\\

\noindent
% Results from the Friedman test between task difficulty and expert annotations
% \begin{table}[]
% \centering
% \caption{\textbf{Results from the Friedman test} between task difficulty and expert annotations. Statistically significant differences are highlighted.}
\label{tab:difficulty_vs_annotations_friedman}
\begin{tabular}{@{}llll@{}}
\toprule
\textbf{Annotation} & \textbf{$\chi^2(2)$} & \textbf{\textit{p}} & \textbf{Kendall's \textit{W}} \\ \midrule
\rowcolor[HTML]{DAE8FC} 
\% on-track & 7.429 & 0.0244 & 0.190 \\
\% stuck & 0.200 & 0.9048 & 0.005 \\
\rowcolor[HTML]{DAE8FC} 
\% uncertain/exploring & 8.143 & 0.0171 & 0.209 \\ \bottomrule
\end{tabular}
% \end{table}

\clearpage

\noindent
\textbf{Results from the Post-hoc Pairwise Wilcoxon test} between task difficulties and self-assessments. Statistically significant pairwise relations are highlighted.\\

\noindent
% Results from the Post-hoc Pairwise Wilcoxon test between task difficulties and self-assessments
% \begin{table}[]
% \centering
% \caption{\textbf{Results from the Post-hoc Pairwise Wilcoxon test} between task difficulties and self-assessments.
% Statistically significant pairwise relations are highlighted.}
\label{tab:difficulty_vs_self-assessments_wilcoxon}
% \resizebox{\columnwidth}{!}{%
\begin{tabular}{@{}llrcc@{}}
\toprule
\textbf{Self-assessment} & \textbf{Pair} & \textbf{W} & \textbf{p} & \textbf{$r_{rb}$} \\ \midrule
Confidence rating & Low $\leftrightarrow$ Medium & 30 & 0.2613 & 0.719 \\
\rowcolor[HTML]{DAE8FC} 
Confidence rating & Low $\leftrightarrow$ High & 4 & 0.0037 & 0.962 \\
\rowcolor[HTML]{DAE8FC} 
Confidence rating & Medium $\leftrightarrow$ High & 7 & 0.0071 & 0.933 \\
NASA\_TLX\_mental & Low $\leftrightarrow$ Medium & 38 & 0.9017 & 0.643 \\
\rowcolor[HTML]{DAE8FC} 
NASA\_TLX\_mental & Low $\leftrightarrow$ High & 6 & 0.0128 & 0.948 \\
\rowcolor[HTML]{DAE8FC} 
NASA\_TLX\_mental & Medium $\leftrightarrow$ High & 5 & 0.0071 & 0.952 \\
NASA\_TLX\_effort & Low $\leftrightarrow$ Medium & 16 & 0.7193 & 0.852 \\
\rowcolor[HTML]{DAE8FC} 
NASA\_TLX\_effort & Low $\leftrightarrow$ High & 0 & 0.0045 & 1.000 \\
\rowcolor[HTML]{DAE8FC} 
NASA\_TLX\_effort & Medium $\leftrightarrow$ High & 0 & 0.0019 & 1.000 \\
NASA\_TLX\_frustration & Low $\leftrightarrow$ Medium & 6 & 0.3173 & 0.943 \\
\rowcolor[HTML]{DAE8FC} 
NASA\_TLX\_frustration & Low $\leftrightarrow$ High & 7 & 0.0320 & 0.933 \\
\rowcolor[HTML]{DAE8FC} 
NASA\_TLX\_frustration & Medium $\leftrightarrow$ High & 8 & 0.0128 & 0.924 \\
NASA\_TLX\_performance & Low $\leftrightarrow$ Medium & 25 & 0.7815 & 0.762 \\
\rowcolor[HTML]{DAE8FC} 
NASA\_TLX\_performance & Low $\leftrightarrow$ High & 4 & 0.0128 & 0.962 \\
\rowcolor[HTML]{DAE8FC} 
NASA\_TLX\_performance & Medium $\leftrightarrow$ High & 0 & 0.0169 & 1.000 \\
NASA\_TLX\_temporal & Low $\leftrightarrow$ Medium & 2 & 0.1573 & 0.976 \\
NASA\_TLX\_temporal & Low $\leftrightarrow$ High & 21 & 0.4902 & 0.800 \\
NASA\_TLX\_temporal & Medium $\leftrightarrow$ High & 9 & 0.1003 & 0.914 \\
NASA\_TLX\_physical & Low $\leftrightarrow$ Medium & 2 & 0.5637 & 0.981 \\
NASA\_TLX\_physical & Low $\leftrightarrow$ High & 0 & 0.0656 & 1.000 \\
NASA\_TLX\_physical & Medium $\leftrightarrow$ High & 0 & 0.0656 & 1.000 \\ \bottomrule
\end{tabular}%
% }
% \end{table}

\vspace{2cm}

\noindent
\textbf{Results from the Post-hoc Pairwise Wilcoxon test} between task difficulties and annotations. Statistically significant pairwise relations are highlighted.\\

\noindent
% Post-hoc Pairwise Wilcoxon test between task difficulties and annotations
% \begin{table}[]
% \centering
% \caption{\textbf{Results from the Post-hoc Pairwise Wilcoxon test} between task difficulties and annotations.
% Statistically significant pairwise relations are highlighted.}
\label{tab:difficulty_vs_annotations_wilcoxon}
% \resizebox{\columnwidth}{!}{%
\begin{tabular}{@{}llrcc@{}}
\toprule
\textbf{Outcome} & \textbf{Pair} & \textbf{W} & \textbf{p} & \textbf{$r_{rb}$} \\ \midrule
\rowcolor[HTML]{DAE8FC} 
\% on-track & Low $\leftrightarrow$ Medium & 8 & 0.0031 & 0.924 \\
\% on-track & Low $\leftrightarrow$ High & 24 & 0.0785 & 0.771 \\
\% on-track & Medium $\leftrightarrow$ High & 27 & 0.1189 & 0.743 \\
\% stuck & Low $\leftrightarrow$ Medium & 10 & 0.4990 & 0.905 \\
\% stuck & Low $\leftrightarrow$ High & 10 & 0.4990 & 0.905 \\
\% stuck & Medium $\leftrightarrow$ High & 10 & 0.2626 & 0.905 \\
\rowcolor[HTML]{DAE8FC} 
\% uncertain/exploring & Low $\leftrightarrow$ Medium & 8 & 0.0031 & 0.924 \\
\rowcolor[HTML]{DAE8FC}
\% uncertain/exploring & Low $\leftrightarrow$ High & 20 & 0.0419 & 0.810 \\
\% uncertain/exploring & Medium $\leftrightarrow$ High & 34 & 0.2676 & 0.676 \\ \bottomrule
\end{tabular}%
% }
% \end{table}

\clearpage

\subsection{Machine learning experiments supplemental results}

\subsubsection{Feature selection matrix}

\textbf{Feature Statistics and Selection Matrix} from the machine learning experiments. Included features are highlighed.\\

\noindent
% \begin{table*}[]
%     % \small
%     \centering
%     \caption{\textbf{Feature Statistics and Selection Matrix} for the machine learning experiments.}
    \label{tab:ML_feature_selections}
    \begin{tabular}{l c c c c c l}
    \toprule
    \textbf{Feature Name} & \textbf{$|d|$} & \textbf{$p$-value} & \textbf{AUC} & \textbf{Var Ratio} & \textbf{Clean} & \textbf{Reason for Exclusion} \\
    \midrule
    \rowcolor[HTML]{DAE8FC} 
    blink\_rate & 0.490 & $4.80 \times 10^{-13}$ & 0.391 & 0.465 & Yes & \\
    \rowcolor[HTML]{DAE8FC} 
    fixation\_dur\_std & 0.484 & $9.05 \times 10^{-13}$ & 0.641 & 0.262 & Yes & \\
    \rowcolor[HTML]{DAE8FC} 
    fixation\_dur\_max & 0.448 & $3.31 \times 10^{-11}$ & 0.636 & 0.238 & Yes & \\
    \rowcolor[HTML]{DAE8FC} 
    fixation\_dur\_mean & 0.420 & $4.90 \times 10^{-10}$ & 0.650 & 0.209 & Yes & \\
    \rowcolor[HTML]{DAE8FC} 
    gaze\_spatial\_range & 0.417 & $6.29 \times 10^{-10}$ & 0.375 & 0.167 & Yes & \\
    \rowcolor[HTML]{DAE8FC} 
    saccade\_count & 0.369 & $4.10 \times 10^{-8}$ & 0.392 & 0.238 & Yes & \\
    fixation\_count & 0.366 & $5.20 \times 10^{-8}$ & 0.395 & 0.266 & No & Multicollinearity ($r=0.95$ with saccade\_count) \\
    \rowcolor[HTML]{DAE8FC} 
    gaze\_vel\_mean & 0.365 & $6.25 \times 10^{-8}$ & 0.352 & 0.217 & Yes & \\
    \rowcolor[HTML]{DAE8FC} 
    head\_pitch\_std & 0.334 & $6.69 \times 10^{-7}$ & 0.393 & 0.324 & Yes & \\
    \rowcolor[HTML]{DAE8FC} 
    mouse\_idle\_ratio & 0.318 & $2.16 \times 10^{-6}$ & 0.397 & 0.182 & Yes & \\
    \rowcolor[HTML]{DAE8FC} 
    head\_roll\_std & 0.302 & $6.85 \times 10^{-6}$ & 0.411 & 0.288 & Yes & \\
    \rowcolor[HTML]{DAE8FC} 
    head\_yaw\_std & 0.284 & $2.37 \times 10^{-5}$ & 0.391 & 0.355 & Yes & \\
    \rowcolor[HTML]{DAE8FC} 
    mouse\_distance & 0.283 & $2.45 \times 10^{-5}$ & 0.626 & 0.322 & Yes & \\
    hr\_max & 0.280 & $2.95 \times 10^{-5}$ & 0.410 & 0.738 & No & Subject-Identity Leak ($>50\%$ variance ratio) \\
    \rowcolor[HTML]{DAE8FC} 
    hr\_std & 0.267 & $6.77 \times 10^{-5}$ & 0.441 & 0.210 & Yes & \\
    \rowcolor[HTML]{DAE8FC} 
    gaze\_vel\_std & 0.251 & $1.79 \times 10^{-4}$ & 0.409 & 0.221 & Yes & \\
    hr\_mean & 0.245 & $2.50 \times 10^{-4}$ & 0.420 & 0.772 & No & Subject-Identity Leak ($>50\%$ variance ratio) \\
    head\_pitch\_mean & 0.236 & $4.11 \times 10^{-4}$ & 0.437 & 0.416 & No & Absolute posture baseline \\
    gsr\_mean & 0.221 & $9.72 \times 10^{-4}$ & 0.556 & 0.908 & No & Subject-Identity Leak ($>50\%$ variance ratio) \\
    hr\_min & 0.212 & $1.56 \times 10^{-3}$ & 0.430 & 0.803 & No & Subject-Identity Leak ($>50\%$ variance ratio) \\
    saccade\_amp\_max & 0.155 & 0.032 & 0.461 & 0.236 & No & Low effect size ($|d| < 0.20$) \\
    gaze\_vel\_max & 0.147 & 0.028 & 0.458 & 0.074 & No & Low effect size ($|d| < 0.20$) \\
    gsr\_slope & 0.066 & 0.322 & 0.526 & 0.057 & No & Not significant ($p > 0.05$) \\
    hrv\_rmssd & 0.065 & 0.332 & 0.529 & 0.382 & No & Not significant ($p > 0.05$) \\
    head\_roll\_mean & 0.058 & 0.381 & 0.511 & 0.057 & No & Absolute posture baseline \& Not significant \\
    saccade\_peak\_vel\_mean & 0.057 & 0.407 & 0.483 & 0.218 & No & Not significant ($p > 0.05$) \\
    saccade\_dur\_mean & 0.055 & 0.409 & 0.491 & 0.322 & No & Not significant ($p > 0.05$) \\
    hrv\_pnn50 & 0.048 & 0.475 & 0.514 & 0.371 & No & Not significant ($p > 0.05$) \\
    gsr\_std & 0.024 & 0.720 & 0.471 & 0.633 & No & Subject Leak ($>50\%$) \& Not significant \\
    gsr\_max & 0.023 & 0.730 & 0.505 & 0.857 & No & Subject Leak ($>50\%$) \& Not significant \\
    saccade\_amp\_mean & 0.022 & 0.761 & 0.509 & 0.211 & No & Not significant ($p > 0.05$) \\
    head\_yaw\_mean & 0.016 & 0.807 & 0.483 & 0.210 & No & Absolute posture baseline \& Not significant \\
    fixation\_disp\_mean & 0.012 & 0.856 & 0.500 & 0.222 & No & Not significant ($p > 0.05$) \\
    saccade\_dir\_std & 0.012 & 0.861 & 0.491 & 0.075 & No & Not significant ($p > 0.05$) \\
    fixation\_disp\_max & 0.007 & 0.923 & 0.508 & 0.311 & No & Not significant ($p > 0.05$) \\
    \bottomrule
    \end{tabular}

    % \end{table*}

\vspace{2cm}

\noindent
\textbf{Model Performance of the 3 Class Classification} ranked by $F1$ score.

\noindent
\label{tab:3_class_ML_ranks}
\begin{tabular}{llcccc}\toprule
    \textbf{Model} & \textbf{Type} & \textbf{Acc.} & \textbf{Precision} & \textbf{Recall} & \textbf{$F1$}\\\midrule
    CNN--LSTM          & Time series  & 0.441 & 0.455 & 0.441 & 0.430\\
    Random Forest      & Tabular      & 0.429 & 0.466 & 0.429 & 0.430\\
    Gradient Boosting  & Tabular      & 0.422 & 0.428 & 0.422 & 0.415\\
    SVM RBF            & Tabular      & 0.423 & 0.417 & 0.423 & 0.411\\
    LSTM               & Time series  & 0.423 & 0.423 & 0.423 & 0.406\\
    2D CNN             & Spectrograms & 0.419 & 0.406 & 0.419 & 0.393\\
    1D CNN             & Time series  & 0.427 & 0.404 & 0.427 & 0.391\\
    MLP                & Tabular      & 0.406 & 0.411 & 0.406 & 0.388\\
    KNN                & Tabular      & 0.402 & 0.395 & 0.402 & 0.377\\
    Decision Tree      & Tabular      & 0.364 & 0.354 & 0.364 & 0.349\\\bottomrule
\end{tabular}

\end{document}